\documentclass[a4paper,11pt]{article}
\usepackage{jheppub}
\usepackage{natbib}
\usepackage{graphicx}
\usepackage{bm}
\usepackage{amssymb}
\usepackage{color}
\usepackage{amsmath}
\usepackage{mathrsfs} 
\usepackage{amstext}
\usepackage[english]{babel}
\usepackage{latexsym}
\usepackage{array}             
\usepackage{multirow}         
\usepackage[usenames,dvipsnames]{xcolor}
\usepackage{soul}

\title{Bubble dynamics and vortex formation in holographic first-order superfluid phase transitions}
\author[a,b]{Zhen-Han Jin}
\author[a,c]{Yu-Ping An}%
\author[a,b,d]{Li Li}
 \affiliation[a]{Institute of Theoretical Physics,
Chinese Academy of Sciences, Beijing 100190, China}
\affiliation[b]{School of Physical Sciences, University of Chinese Academy of Sciences, Beijing 100049, China}
\affiliation[c]{Department of Physics, Technion, Haifa 32000, Israel}
\affiliation[d]{School of Fundamental Physics and Mathematical Sciences, Hangzhou Institute for Advanced Study, University of Chinese Academy of Sciences, Hangzhou 310024, China}
\emailAdd{jinzhenhan@itp.ac.cn, yuping.an@campus.technion.ac.il, liliphy@itp.ac.cn}

\abstract{
We investigate bubble dynamics in a holographic superfluid undergoing a first‑order phase transition with spontaneous $U(1)$ symmetry breaking. Near the nucleation threshold, the system exhibits universal critical behavior governed by a single unstable mode, leading to logarithmic scaling of the time spent near the critical solution. The terminal bubble wall velocity increases with charge density but remains small due to strong dissipation. In multi‑bubble collisions, vortex formation depends sensitively on the initial phases and deviates significantly from the geodesic rule. Notably, we identify a regime where three‑bubble collisions produce a vortex‑antivortex pair that subsequently annihilates, a phenomenon not predicted by the geodesic rule. The lifetime of this pair scales logarithmically with the distance to the critical collision radius. Our results underscore the crucial role of non‑equilibrium dynamics in strongly coupled superfluids and provide new insights into topological defect formation during first‑order phase transitions.

}
\vspace{1cm}

\begin{document}

\maketitle

\section{Background}

First-order phase transitions, characterized by metastable states and discontinuous order parameter changes, can proceed via bubble nucleation. This fundamental process has been extensively studied across a wide range of physical systems, from liquid-gas systems~\cite{Asano:2022} and quantum materials~\cite{Sinha:2021} to early universe cosmology~\cite{Davis:1999ii, Dine:1992wr}. Across these diverse settings, bubble dynamics plays a crucial role in determining how a first-order transition unfolds and in producing observable signatures. In cosmology, for example, the dynamics of expanding and colliding bubbles is an important source of gravitational waves~\cite{Jinno:2016vai,Kim:2014ara}. It can also drive baryogenesis~\cite{Baldes:2021vyz}, lead to dark matter production~\cite{Azatov:2021ifm}, and contribute to primordial black hole formation~\cite{Jung:2021mku}. Quantitative features such as the bubble wall velocity significantly influence gravitational wave spectra~\cite{Gowling:2021gcy}, motivating extensive investigations~\cite{Branchina:2025adj,Yuwen:2024hme,Li:2023xto}. Beyond cosmological settings, molecular dynamics studies of bubble nucleation and dynamics have also gained attention in condensed matter and fluid systems, aiming to uncover the nanoscale mechanisms governing phase transitions and heat transfer~\cite{Fallahzadeh:2024,Lin:2024}.

Bubble nucleation and growth are intrinsically non-equilibrium processes, often involving strong coupling and large fluctuations that defy perturbative treatments. Holographic duality offers a powerful framework for investigating such systems by mapping the non-equilibrium dynamics of a strongly coupled field theory to the evolution of a dual classical gravity background, which can be efficiently studied through numerical simulations. This approach has been successfully applied to first-order phase transitions, providing insights into several key aspects. For example, it has been used to study dynamical stability in holographic phase transitions~\cite{Zhao:2022jvs,Zhao:2023ffs,Chen:2022cwi,Janik:2017ykj,Janik:2015iry,Attems:2020qkg,Bellantuono:2019wbn} and non-equilibrium dynamics in $Z_2$-symmetry-breaking processes~\cite{Bea:2021zsu,Bea:2022mfb,Zhao:2026eav}. In particular, the critical behavior during first-order phase transitions was examined in~\cite{Chen:2022cwi}, where the scaling law and stability of the critical solution were analyzed. This behavior is reminiscent of black hole critical collapse~\cite{Liebling:1996dx,Choptuik:1996yg,Choptuik:2004ha} as well as the first-order phase transition of a two-dimensional $\phi^4$ model~\cite{Munster:2000}. Despite these advances, however, several important aspects of bubble dynamics remain largely unexplored. These include multi-bubble interactions, the effects of bubble collisions, and critical nucleation phenomena in holographic superfluids with $U(1)$ symmetry breaking. Understanding these processes is essential for connecting holographic models to realistic phase transitions in both cosmological and condensed matter settings.

A distinctive feature of bubble dynamics in systems with $U(1)$ symmetry breaking, compared to the $Z_2$ case, is the formation of topological defects—specifically, vortices. In second-order phase transitions, the density of such vortices can be accurately described by the Kibble--Zurek mechanism (KZM)~\cite{Kibble:1976sj,Zurek:1985qw,Zurek:1996sj}, which has found broad applications across condensed matter physics, cosmology, and quantum computation~\cite{Ruutu:1996,Maniv:2003,Keesling:2018ish}. However, the KZM is not directly applicable to first-order phase transitions, where bubble nucleation and expansion dominate the dynamics. Recently, Zurek and collaborators attempted to extend the KZM to phase transitions with tunable order by combining the standard KZM scaling with nucleation theory, thereby predicting defect densities~\cite{Suzuki:2023hag}. A deeper understanding of defect formation during first-order transitions requires careful examination of the so-called geodesic rule. The validity and applicability of this rule remains an open question, and addressing it is crucial for establishing a unified framework for topological defect formation across different types of phase transitions.


The geodesic rule~\cite{Kibble:1976sj} assumes that during domain merging, order parameters follow the shortest path in the vacuum manifold to minimize gradient energy. For $U(1)$ symmetry breaking, this predicts a vortex formation probability of $1/4$ in three-bubble collisions when phases are randomly assigned. This rule has found support in various systems including nematic liquid crystals~\cite{Bowick:1992rz} and Bose-Einstein condensates~\cite{Aidelsburger:2017dnn}. Deviations from the geodesic rule have also been reported. Early numerical simulations revealed vortex-antivortex pair production during bubble collisions~\cite{Srivastava:1991nv}, which were attributed to large field oscillations that cause the field overshooting of the potential barrier~\cite{Digal:1996ip}. Subsequent studies showed that defect formation can also be affected by phase relaxation dynamics~\cite{Kibble:1995aa,Copeland:1999ua}, bubble size, expansion velocity, interaction strength~\cite{Digal:1995vd,Digal:1996ip,Digal:1997gc}, and damping effects from the surrounding environment~\cite{Ferrera:1995ef,Ferrera:1996hu}. As a self-consistent theory, which naturally incorporates strong coupling and finite temperature effects, holography offers a unique platform to examine validity of geodesic rule in these circumstances.

This paper extends previous holographic studies by systematically investigating bubble dynamics across three key stages: nucleation, expansion, and multi‑bubble collisions. Using a holographic model with broken $U(1)$ symmetry, we provide a comprehensive analysis that reveals critical phenomena and vortex formation processes, offering deeper insight into the geodesic rule. The remainder of this paper is organized as follows. Section~\ref{Holographic Setup} describes the holographic framework employed in our study. Section~\ref{nucleation and critical} examines the nucleation process, with a particular focus on critical behavior near the bubble nucleation threshold. Section~\ref{bubble_wall_v} is then devoted to the wall velocity of an expanding superfluid bubble. Section~\ref{defect_generation} addresses vortex formation resulting from multi-bubble collisions. Section~\ref{conclusion} summarizes our main findings. Finally, technical details are provided in Appendices~\ref{app:A} and~\ref{app:B}.

\section{Holographic Setup}
\label{Holographic Setup}
We study a holographic superfluid model with nonlinear self-interactions in a (3+1)-dimensional asymptotically anti-de Sitter (AdS) spacetime. The bulk action is given by:
\begin{equation}
S = \int d^{4}x \sqrt{-g} \left[ \frac{1}{2\kappa_N^2}\left(R + \frac{6}{L^{2}} \right) + \mathcal{L}_m \right]\,,
\end{equation}
where $L$ is the AdS radius and the matter Lagrangian $\mathcal{L}_m$ reads~\cite{Zhao:2022jvs}

\begin{equation}
\mathcal{L}_m = -\frac{1}{4}F^{\mu\nu}F_{\mu\nu} - \left( |D_\mu\Psi|^{2} - m^{2}|\Psi|^{2} - \lambda (\Psi^*\Psi)^2 - \tau (\Psi^*\Psi)^3 \right)\,.
\end{equation}
Here, $D_{\mu} = \nabla_\mu - iq A_\mu$ denotes the covariant derivative, $A_\mu$ is a $U(1)$ gauge field with field strength $F_{\mu\nu} = \partial_\mu A_\nu - \partial_\nu A_\mu$, and $\Psi$ is a complex scalar field. The parameters $\lambda$ and $\tau$ control the strength of the self-interactions, with which first-order, second order and zeroth order phase transitions can be realized~\cite{Zhao:2022jvs}.

For simplicity, we work in the probe limit, where the back-reaction of the matter fields onto the geometry is neglected. This approximation effectively decouples the dynamics of the geometrical sector from the matter sector, freezing the fluctuations of the temperature and the normal fluid's velocity. In this limit, the background geometry is fixed as an AdS-Schwarzschild black brane:
\begin{equation}\label{AdSsw}
ds^2 = \frac{L^2}{z^2} \left( -f(z) dt^2 - 2 dt dz + dx^2 + dy^2 \right)\,,
\end{equation}
where $f(z) = 1 - (z/z_h)^3$ with $z_h$ the location of the horizon. The Hawking temperature of this black brane, which corresponds to the temperature of the boundary heat bath, is $T = \frac{3}{4\pi z_h}$.
Henceforth, we set $L=1$, $m^2 = -2$ and $q=1$ for simplicity. We also adopt the radial gauge $A_z = 0$. The dynamics of the matter fields are simulated by integrating their equations of motion (the details of the time evolution are provided in Appendix~\ref{app:A}).

According to the holographic dictionary, physical observables of the boundary quantum system can be extracted from coefficients of the asymptotic expansions of the bulk fields near the AdS boundary ($z \to 0$):
\begin{align}
A_\mu = a_\mu + b_\mu z + ...\,, \;\;
\Psi = \psi^{(s)} z + \psi^{(v)} z^2 + ...\,.
\end{align}
Here $\psi^{(s)}$ is the source term. To realize spontaneous $U(1)$ symmetry breaking, we set this source to zero, $\psi^{(s)} = 0$. Thus, the subleading coefficient $\psi^{(v)}$ corresponds to the superfluid condensate $\mathcal{O}=|\mathcal{O}|e^{i\theta}$. The spatial components of the leading term of the gauge field, together with the gradient of the condensate phase, give the superfluid velocity $v^s_i = \partial_i \theta - a_i$ with $i = x, y$. We set $a_i = 0$ in practice. The charge density is read off from the subleading term in the near-boundary expansion of the gauge field:
\begin{equation}
    b_t = -\rho\,.
\end{equation}
Finally, there remains a residual $U(1)$ symmetry
\begin{equation}\label{residual}
 A_t \rightarrow A_t + \partial_t \Sigma(t)\,,\quad \Psi \rightarrow e^{iq\Sigma(t)}\, \Psi \,,  
\end{equation}
where $\Sigma(t)$ depends only on $t$. The chemical potential $\mu$, which is invariant under this residual symmetry, is then defined as
\begin{equation}
    \mu = A_t\big|_{z=0} - A_t\big|_{z=z_h}\,,
\end{equation}
measuring the difference of the gauge potential $A_t$ between the boundary and the horizon. Note that the superfluid velocity $v^s_i$ and the charge density $\rho$ remain unchanged under the residual symmetry~\eqref{residual}.

The system possesses the following scaling symmetry:
\begin{equation}
(t, x, y, z) \to \xi (t, x, y, z)\,, \; z_h \to \xi z_h\,, \; (T, \mu) \to \xi^{-1}(T, \mu),\, \; (\rho, \mathcal{O}) \to \xi^{-2}(\rho, \mathcal{O})\,,
\end{equation}
where $\xi$ is a constant. We set $z_h = 1$ (\emph{i.e.}, $T = \frac{3}{4\pi}$) throughout this work. Therefore, $\rho$ is the only free parameter for fixed theory parameters. Equivalently, all physical results depend solely on the dimensionless ratio $\rho/T^2$. The phase structure is explored by varying the charge density $\rho$, which corresponds to working in the canonical ensemble. Moreover, a physical time $t$ should be combined with the temperature to form the dimensionless product $tT$. However, for notational simplicity and without loss of generality, we will omit the explicit factor of $T$ when discussing time evolution in the following sections.

\begin{figure}[htbp]
\centering
\includegraphics[width=0.6\textwidth]{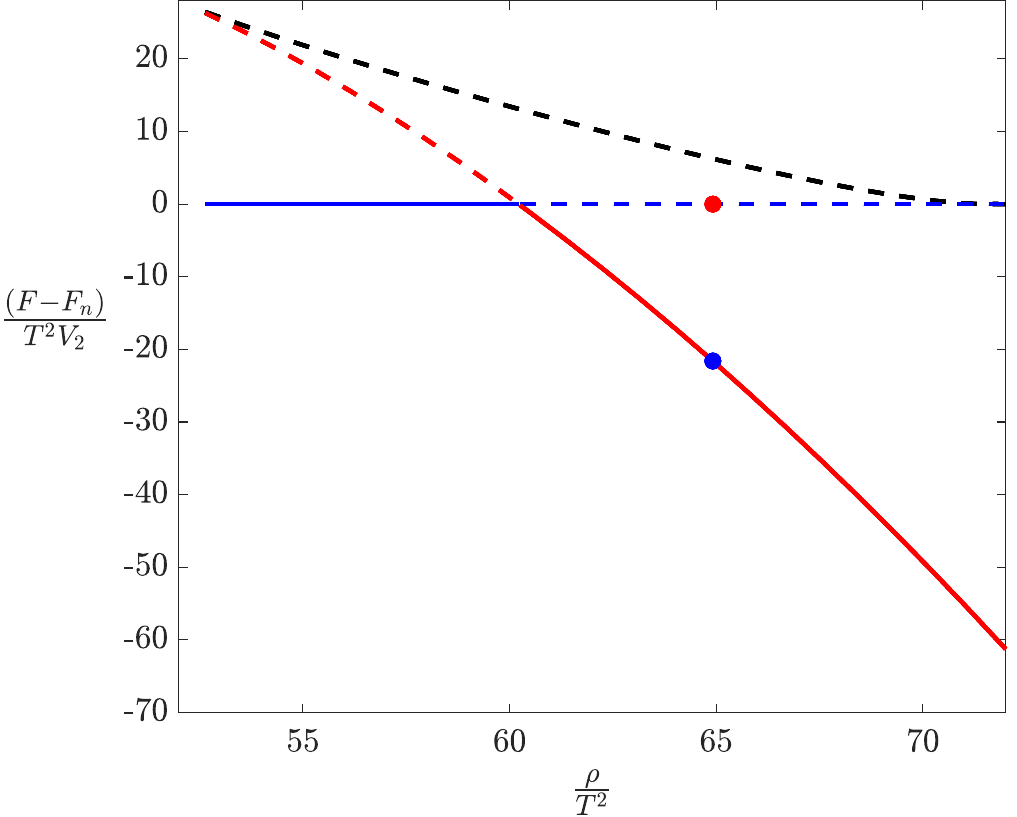}
\caption{Free energy density$\frac{(F-F_n)}{T^2V_2}$ as a function of charge density$\frac{\rho}{T^2}$ for the static homogeneous solutions with parameter set as $\lambda = -2,\tau = 0.8$. The normal state is shown in blue and the superfluid state in red. Solid lines (both blue and red) represent stable states. Dashed red and blue lines correspond to metastable states, while the dashed black line indicates a state that is both thermodynamically and dynamically unstable. Our dynamical simulation starts from the metastable normal state, indicated by the red point. By applying a sufficiently large perturbation, the system evolves into the superfluid state, represented by the blue point.}
\label{F_rho}
\end{figure}

The relevant thermodynamic potential in the canonical ensemble is the free energy $F$. According to the holographic dictionary, $F$ is given by the  Euclidean on-shell action under appropriate  boundary terms~\cite{Hartnoll:2009sz}. In the probe limit, where the gravitational background is fixed, it suffices to consider the free energy of matter fields. For homogeneous and static solutions, it takes the form:
\begin{equation}
F = \frac{V_2 L}{T} \left[ \frac{\mu \rho}{2} + \int_0^{z_h} \mathrm{d}z \left( \frac{A_t^2 \psi^2}{f} - \lambda \psi^4 - 2 \tau \psi^6 \right) \right]\,,
\label{free_energy_cal}
\end{equation}
where $V_2$ is the spatial volume of the boundary system. As shown in~\cite{Zhao:2022jvs}, the type of superfluid phase transition depends on the parameters $\lambda$ and $\tau$. In this work, we choose $\lambda$ and $\tau$ from a parameter region that gives rise to a first-order transition. Using~\eqref{free_energy_cal}, we plot the free energy density of the static homogeneous solutions as a function of charge density in Fig.~\ref{F_rho}. Here, $F_n$ denotes the free energy of the normal fluid. The normal state is shown in blue and the superfluid state in red. Solid lines (both blue and red) represent thermodynamically stable states. The discontinuous first derivative of the free energy with respect to the charge density signals a first-order phase transition. The dashed red and blue lines correspond to metastable states, while the dashed black line indicates a state that is both thermodynamically and dynamically unstable. Our dynamical simulation begins from the metastable normal state indicated by the red point. Upon applying a sufficiently large perturbation, the system undergoes a phase transition and evolves to the stable superfluid state (represented by the blue point) through the nucleation of superfluid bubbles.
\\

\section{Bubble nucleation and critical behavior}
\label{nucleation and critical}
We begin by examining single-bubble dynamics during a first-order superfluid phase transition. It is well known that the transition from a metastable state to the globally stable phase is hindered by a potential barrier. To nucleate a bubble, a sufficiently strong perturbation is therefore required to overcome this barrier and trigger the phase transition. Dynamically, if the perturbation drives the system from the metastable state into the globally stable one, the evolution is termed supercritical; if instead the perturbation decays and the system returns to the metastable state, the evolution is subcritical. The solution that separates supercritical from subcritical evolution is known as the critical solution. It corresponds to the minimal dynamical excitation needed to initiate the phase transition and thus represents the dynamical threshold of the system.

Near the critical threshold for bubble nucleation, systems often exhibit universal critical behavior, reminiscent of the critical phenomena observed in gravitational collapse. Previous studies, such as~\cite{Chen:2022cwi}, have explored this behavior in the context of an explicitly broken $Z_2$ symmetry, where the order parameter is not well defined. In that setup, the phase transition drives the system from a metastable state to a phase-separated state, which, from the bulk perspective, corresponds to a transition between two distinct scalarized black hole phases. In contrast, our work investigates the spontaneous breaking of a $U(1)$ symmetry, which describes a transition from a black hole without scalar hair to a scalarized black hole. Consequently, the order parameter in our system—the superfluid condensate—remains well defined throughout the entire transition.
\begin{figure}[htbp]
\centering
\includegraphics[width=0.65\textwidth]{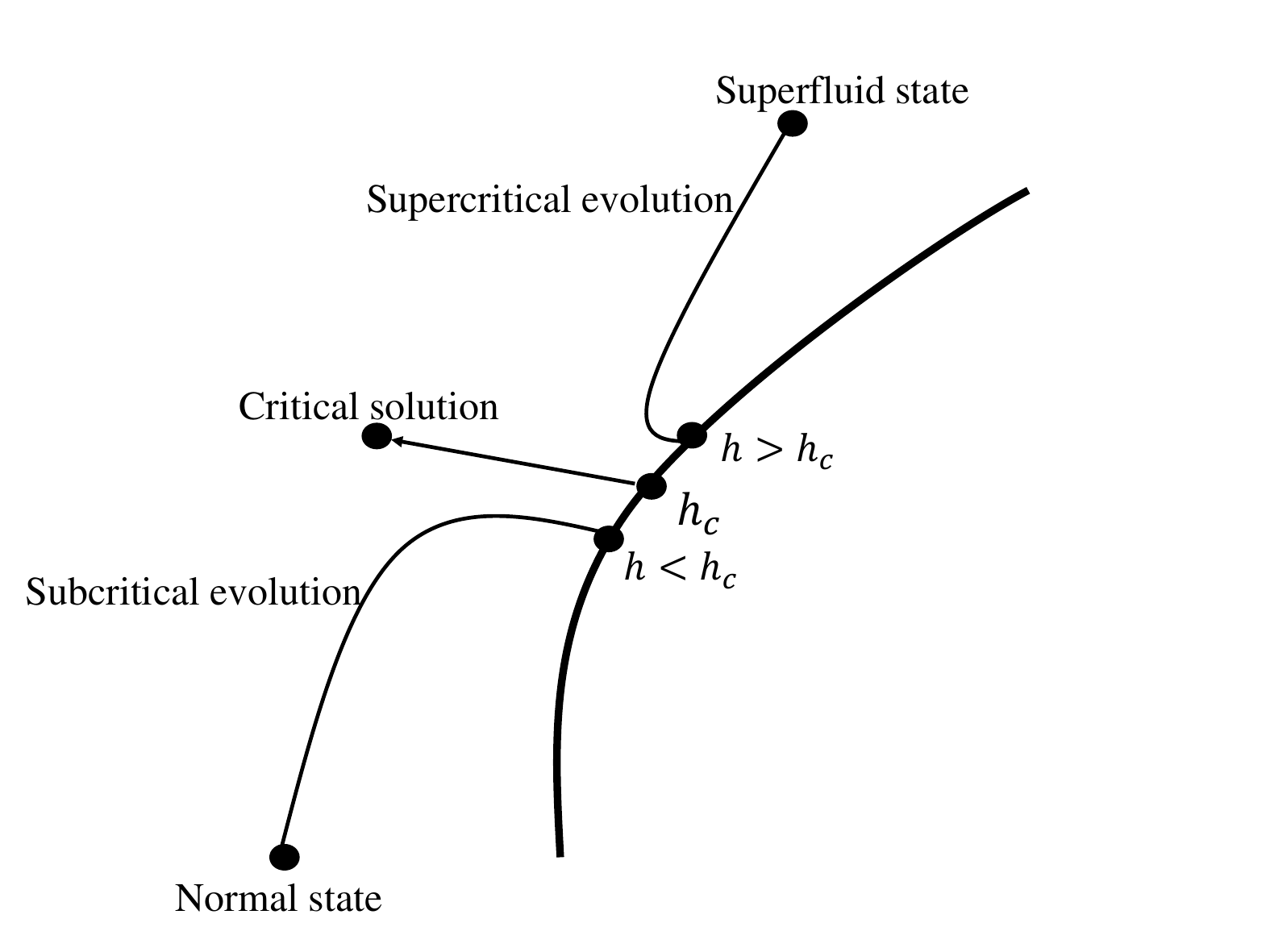}
\caption{Evolution of the system under different initial perturbation amplitudes. When the amplitude equals the critical value $h = h_c$, the system remains at the critical solution. For $h > h_c$, the evolution becomes supercritical, driving the system toward the stable superfluid state. For $h < h_c$, the evolution remains subcritical, and the system returns to the metastable normal state.}
\label{critical_solution}
\end{figure}

Without loss of generality, we employ a Gaussian wave packet as the initial perturbation:
\begin{equation}
\delta\mathcal{O} = h \exp\left(-\frac{r^2}{2\sigma^2}\right)\,,
\label{gaussian_wave}
\end{equation}
where $h$ is the amplitude, $\sigma$ is the width, and $r$ is the radial distance from the nucleation center. In our simulations, the width is fixed at $\sigma=5$, while the amplitude $h$ is varied to probe the dynamics near the critical point. For $h<h_c$, the evolution remains subcritical; for $h>h_c$, it becomes supercritical. The evolution at the precise threshold $h=h_c$ defines the critical solution, illustrated schematically in Fig.~\ref{critical_solution}. Using the bisection method, we numerically determine the critical amplitude to be $h_c\approx1.186469467$. 

\begin{figure}[htbp]
    \centering
        \includegraphics[width=1\textwidth]{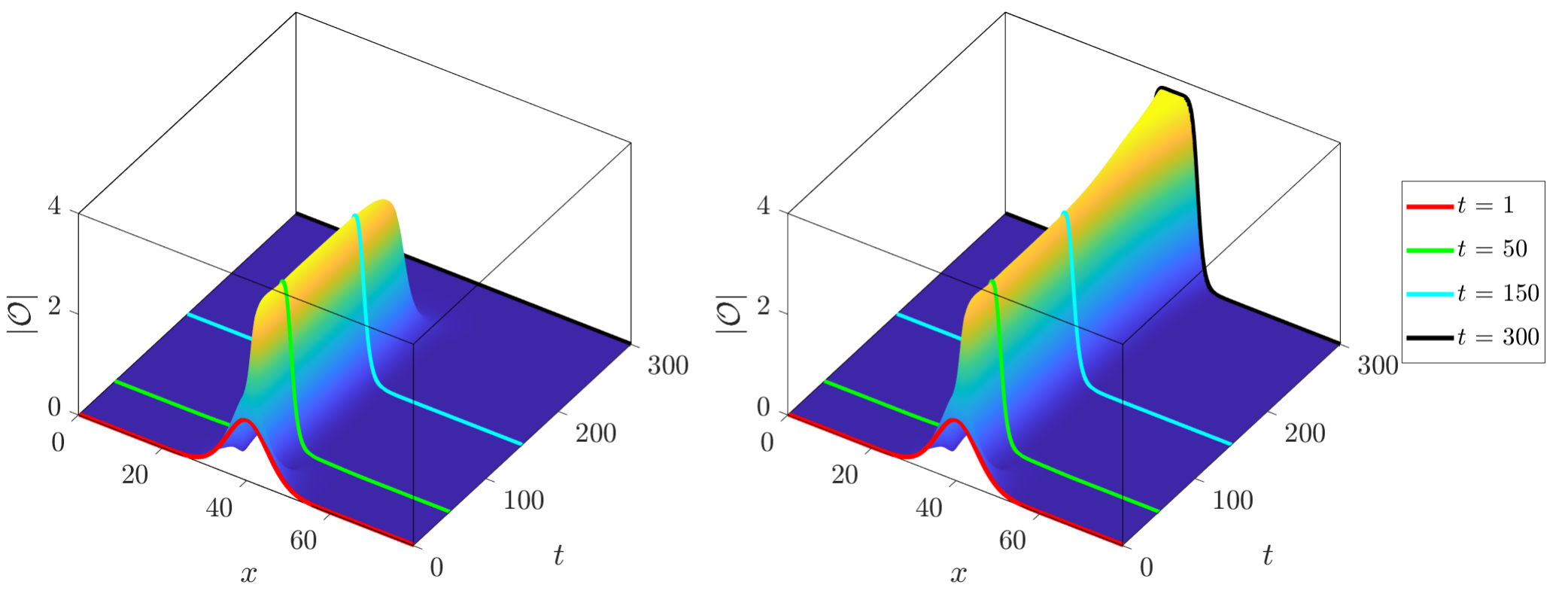}
    \caption{Evolution of the condensate for initial perturbations at $h = h_c \pm e^{-12}$, with $\lambda = -2, \tau = 0.8,\sigma = 5$ and $\rho = 3.7$. \textbf{Left}: subcritical case with $h = h_c - e^{-12}$, where the perturbation decays. \textbf{Right}: supercritical case with $h = h_c + e^{-12}$, where a bubble forms. Profiles at four representative time slices are shown in distinct colors.}
    \label{fig:conden_t}
\end{figure}

\begin{figure}[htbp]
\centering
\includegraphics[width=0.495\textwidth]{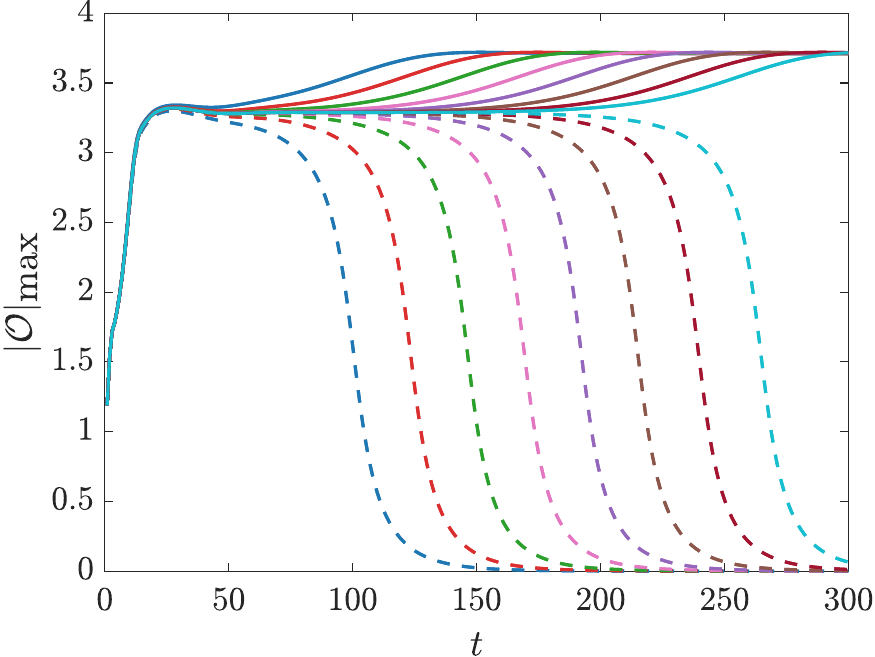}
\includegraphics[width=0.495\textwidth]{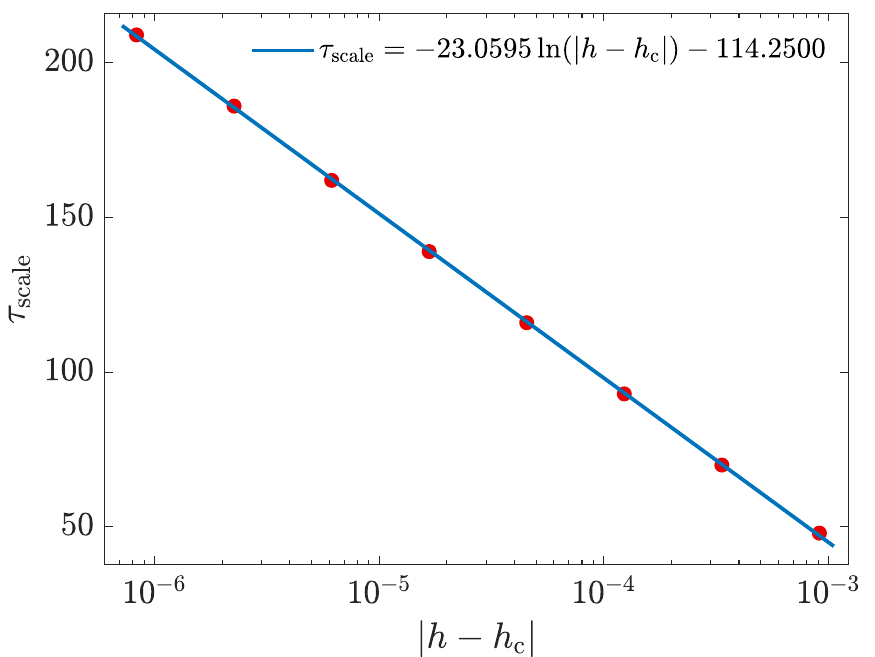}

\caption{Single bubble dynamics near the critical solution, with $\lambda = -2, \tau = 0.8,\sigma = 5$ and $\rho = 3.7$. Left: Time evolution of the maximum condensate value for initial perturbation amplitudes $h = h_c \pm e^{-m}$ with $m \in [7,14]$. Solid and dashed lines correspond to supercritical ($h > h_c$) and subcritical ($h < h_c$) evolutions, respectively. Lines of the same color share the same exponent $m$, representing equal logarithmic distance from the critical point but on opposite sides. Right: Scaling of $\tau_{\mathrm{scale}}$ as a function of $\ln |h - h_c|$. The numerical data (dots) follow the linear scaling relation (blue line) predicted by~\eqref{tau} with $\tau_0 = 23.0595$ .}
\label{fig:total}
\end{figure}

To examine near-critical behavior, we fine-tune the initial amplitude to $h = h_c \pm e^{-12}$, where the plus (minus) sign leads to a supercritical (subcritical) outcome. Exploiting the rotational symmetry of the single-bubble system, Fig.~\ref{fig:conden_t} shows the evolution of a single slice of the condensate with nucleation center at $x=40$. The evolution can be broadly divided into three distinct stages.
\begin{itemize}
    \item In the first stage (before green lines), the Gaussian wave packet narrows and sharpens, as illustrated by the green curve. This initial dynamics is essentially identical for both supercritical and subcritical cases. 
    \item The second stage (between green and cyan lines) is characterized by a slow growth (supercritical) or gradual decay (subcritical) of the wave packet. Crucially, during this stage, the profiles of the two cases remain qualitatively similar, making it difficult to distinguish the eventual fate of the system.
    \item The final stage (after cyan lines) reveals the critical difference: for the supercritical case, a bubble nucleates and expands, while for the subcritical case, the perturbation decays completely, leaving the system in the metastable normal state. This dichotomy clearly illustrates the threshold behavior associated with a first-order phase transition.
\end{itemize}
Moreover, we observe that for initial parameters close to $h_c$, the field configuration is attracted toward the critical solution, though it does not settle into it precisely. The closer $h$
is to $h_c$, the longer the system lingers near the critical solution. The left panel of Fig.~\ref{fig:total} shows the time evolution of the maximum of the
condensate for the initial data of the form \( h = h_c \pm e^{-m} \) with \( m \in [7, 14] \). Dashed lines correspond to subcritical cases, while solid lines of the same color represent supercritical evolutions. It is natural to conjecture that exactly at $h=h_c$, the solution would remain static—corresponding to the critical solution.

The static nature of the critical solution enables a linear perturbation analysis around it. Following the analogy with critical phenomena in gravitational collapse~\cite{Gundlach:2025yje}, we assume the existence of a single unstable mode when the field configuration is sufficiently close to the critical solution: 
\begin{equation}
\mathcal{O}(t, \mathbf{r}) = \mathcal{O}_c(\mathbf{r}) + (h - h_c) e^{ t/\tau_0} \delta\mathcal{O}_0(\mathbf{r}) + \text{(decaying modes)}\,,
\label{eq:perturbation}
\end{equation}
where $\mathbf{r}=(x,y)$ denotes spatial coordinates, $\mathcal{O}_c(\mathbf{r})$ is the critical solution, and $\delta\mathcal{O}_0(\mathbf{r})$ is the eigenmode corresponding to the eigenvalue $1/\tau_0$ of the linearized operator. As time increases, this perturbation grows and eventually becomes non-perturbative. For a fixed deviation
\begin{equation}
\mathcal{O}(t, \mathbf{r}) - \mathcal{O}_c(\mathbf{r}) = (h - h_c) e^{\tau_\text{scale}/\tau_0} \delta\mathcal{O}_0(\mathbf{r})\,,
\label{epsilon_h}
\end{equation}
the time $\tau_\text{scale}$ during which the configuration stays close to the critical solution satisfies
\begin{equation}
\label{tau}
\tau_\text{scale} \sim -\tau_0 \ln |h - h_c|\,.
\end{equation}

To measure the time $\tau_\text{scale}$ from our numerical simulations, we track the maximum condensate value $\mathcal{O}_{\text{max}}(t)$ for a matched pair of supercritical and subcritical evolutions that share the same deviation $|h - h_c|$. The scaling time $\tau_\text{scale}$ is then operationally defined as the instant at which the difference between these two dynamical branches exceeds a chosen threshold $\epsilon$:
\begin{equation}
|\mathcal{O}_{\text{max}}^s(\tau_\text{scale}) - \mathcal{O}_{\text{max}}^n(\tau_\text{scale})| = \epsilon\,,
\end{equation}
where the superscripts $s$ and $n$
label the supercritical and subcritical trajectories, respectively. According to the linear perturbation ansatz of~\eqref{epsilon_h}, this empirically defined $\tau_\text{scale}$ is expected to follow the same logarithmic scaling law given in~\eqref{tau}. To implement this, we set $\epsilon = 0.1$ and measure $\tau_\text{scale}$ across a range of initial amplitudes $h$. The results, displayed in the right panel of Fig.~\ref{fig:total}, clearly confirm the predicted linear relation between $\tau_\text{scale}$ and $\ln |h - h_c|$, thereby validating the presence of a single unstable mode governing the near-critical dynamics. 

\section{Bubble wall velocity}
\label{bubble_wall_v}
The initially nucleated static bubble begins to accelerate under a thermodynamic driving force—typically due to the free energy difference between the metastable and stable phases. In a steady-state expansion, this driving force may eventually be balanced by dissipative backreactions from the surrounding medium. The terminal wall velocity reached during such a bubble expansion in first-order phase transitions is a key dynamical parameter. It plays a crucial role in several important physical phenomena, notably in determining the density and distribution of topological defects formed during the phase transition~\cite{Kibble:1995aa,Borrill:1995gu}, and in setting the peak frequency and amplitude of the stochastic gravitational wave background produced by bubble collisions and sound waves~\cite{Jinno:2016vai,Kim:2014ara}. In this section, we quantitatively investigate the bubble wall velocity within the framework of holographic superfluids. By varying physical conditions such as temperature and charge density, we systematically compute the terminal velocity of an expanding superfluid bubble in a strongly coupled system. This provides quantitative insights into the dynamics of phase interface propagation under strong coupling, offering a valuable benchmark for effective field theories and cosmological phase transition models.

In our numerical simulations, the perturbation amplitude and width are fixed at $h = 1.25$ and $\sigma = 5$. We investigate the expansion dynamics of a single bubble across different charge densities ($T$ is fixed), focusing on the range $\rho \in [3.66, 4]$. This interval is selected based on clear physical constraints: at higher charge densities, the potential barrier between the metastable normal phase and the stable superfluid phase becomes very small, allowing even weak fluctuations to trigger nucleation and making the bubble wall poorly defined. Conversely, at lower charge densities, the barrier is too large for our fixed perturbation ($h=1.25, \sigma=5$) to initiate nucleation reliably. We then define the instantaneous bubble radius $r(t)$ as the radial distance from the bubble center to the contour where the condensate reaches half of its equilibrium value. The bubble wall velocity $v(t)$ is then derived from the temporal derivative of the radius.

\begin{figure}[htbp]
    \centering     \includegraphics[width=0.49\textwidth]{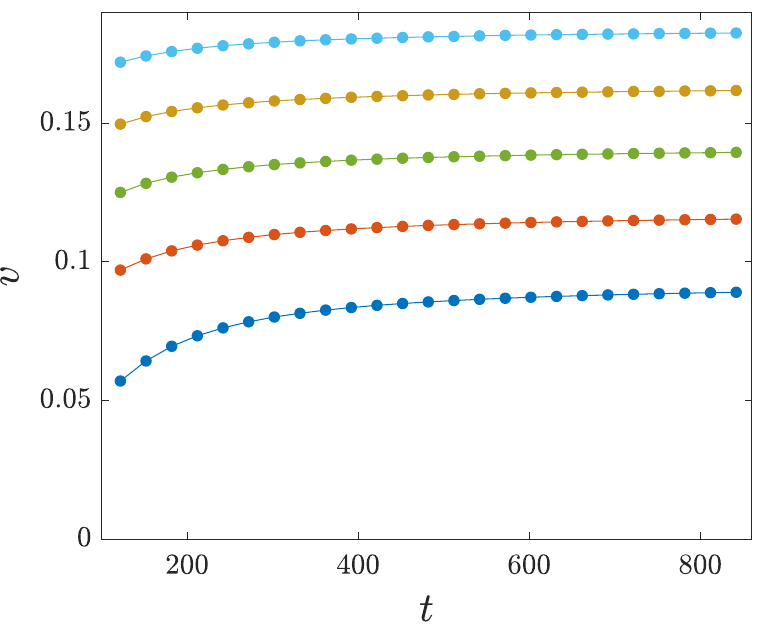}    \includegraphics[width=0.49\textwidth]{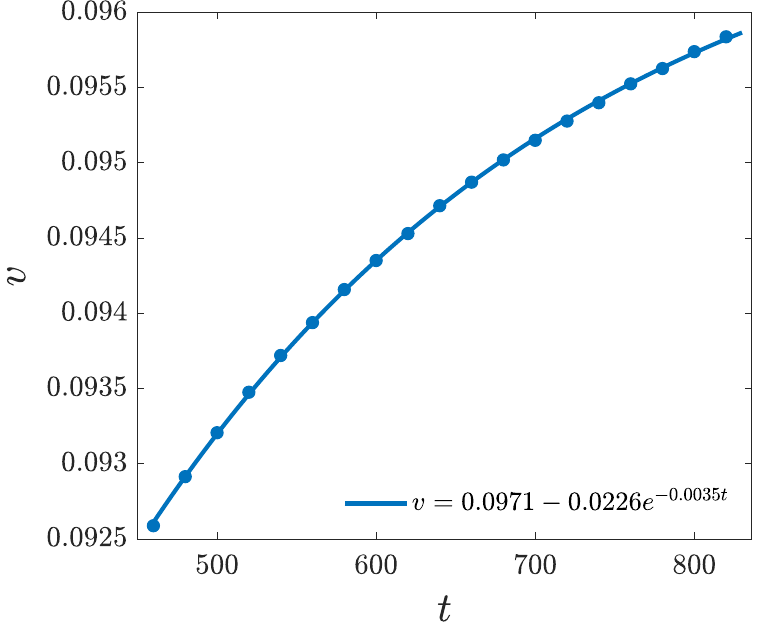}    
    \caption{Time evolution of the bubble wall velocity in holographic superfluid phase transitions, with $\lambda = -2, \tau = 0.8, h = 1.25$ and $\sigma = 5$. The data points come from numerical simulations of the holographic model. \textbf{Left:} Bubble-wall velocity as a function of time for charge densities $\rho \in \{3.68, 3.76, 3.84, 3.92, 4.00\}$; from bottom to top, the curves correspond to increasing $\rho$. \textbf{Right:} Detailed view of the late-time evolution for $\rho = 3.7$.The solid curve shows the fit to the late-time data ($t>460$), using the form $v(t) = a - b e^{-ct}$. Best-fit parameters are $a = 0.0971$, $b = 0.0226$, and $c = 0.0035$, with a high coefficient of determination ($R^2 = 0.999872$).}
    \label{fig:rho_v}
\end{figure}

The right part of Fig.~\ref{fig:rho_v} shows the time evolution of the bubble wall velocity for different charge densities $\rho$. The wall acceleration is found to decrease over time. Although computational constraints prevent extremely long simulations, the observed trend indicates that the velocity asymptotically approaches a terminal value, denoted as $v_{\text{terminal}}$. To characterize this behavior, we fit the late time data ($t>460$) to a function of the form $v_{\text{terminal}} - c e^{-b t}$, where $b$ and $c$ are positive constants. As shown by the blue fitting curves in the right panel of Fig.~\ref{fig:rho_v}, this form provides a good description of the late-time dynamics, with a coefficient of determination $R^2 = 0.999872$ for the case $\rho = 3.7$. This late-time behavior of the bubble wall velocity to a terminal value is consistent with the behavior reported in~\cite{Ferrera:1995ef}.

\begin{figure}[htbp]
\centering
\includegraphics[width=0.55\textwidth]{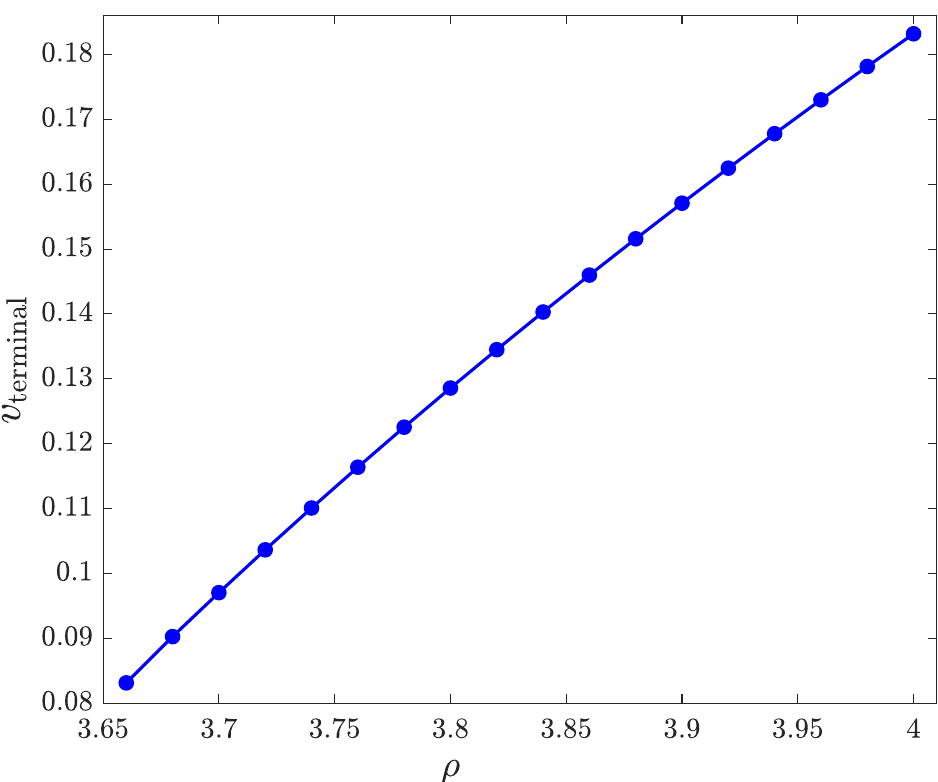}
\caption{Terminal bubble-wall velocity $v_{\text{terminal}}$ as a function of the charge density $\rho$. The terminal velocity increases monotonically with $\rho$. We have set $\lambda = -2, \tau = 0.8,h = 1.25$ and $\sigma = 5$.}
\label{rho_v_scale}
\end{figure}

The dependence of the terminal velocity 
$v_{\text{terminal}}$ on the charge density $\rho$ is summarized in Fig.~\ref{rho_v_scale}, showing a clear monotonic increase. Notably, our holographic simulations yield a non-relativistic terminal wall velocity. This behavior can be understood through the strong coupling between the expanding bubble wall and the surrounding thermal plasma. The backreaction force opposing the wall's motion grows rapidly, balancing the thermodynamic driving force after only a brief period of acceleration. This non-relativistic regime stands in contrast to the relativistic case, where the wall moves so quickly that local thermal equilibrium near the interface cannot be maintained. In our scenario, the wall velocity is sufficiently low to preserve approximate local thermal equilibrium around (though not precisely at) the wall, validating a hydrodynamic description of the plasma as a near-perfect fluid~\cite{Li:2023xto,Wang:2023lam}. This picture is consistent with findings from other strongly coupled holographic models (see, \emph{e.g.},~\cite{Bea:2021zsu, Janik:2022wsx,Bigazzi:2021ucw,Bea:2022mfb}).
The quantitative scaling of $v_{\text{terminal}}$ with $\rho$, combined with its observed asymptotic saturation, provides key insight into the expansion dynamics of first-order phase transitions in strongly coupled, dissipative systems. Furthermore, as discussed in the following section, the bubble wall velocity critically influences subsequent non-equilibrium dynamics, particularly the processes of vortex formation.

\section{Vortex formation in first-order phase transition}
\label{defect_generation}
The formation of vortex during phase transitions is a fundamental problem across diverse physical systems. The KZM provides predictions for vortex densities in second-order transitions by relating them to equilibrium critical exponents~\cite{Kibble:1976sj,Zurek:1985qw,Zurek:1996sj}. However, its direct application to first-order transitions remains limited. While recent work~\cite{Suzuki:2023hag} has extended KZM-type frameworks to first-order scenarios, the dynamics of vortex formation in such systems are still not fully understood. This motivates our focused investigation of the geodesic rule—a key assumption that governs phase interpolation during domain coalescence. In this section, we examine vortex formation during bubble collisions in our holographic superfluid and systematically quantify deviations from the geodesic rule in strongly non-equilibrium environments.

A vortex is a type of topological defect that can arise when the vacuum manifold has nontrivial homotopy. In the case of a spontaneously broken $U(1)$ symmetry considered in this work, the vacuum manifold is a circle $S^1$, and vortex configurations are classified by the first homotopy group $\pi_1(S^1)=\mathbb{Z}$. Specifically, in the presence of a vortex, a closed loop in physical space that is homeomorphic to $S^1$ and encircles the defect core will have the phase of the order parameter winding nontrivially around the vacuum manifold. The phase configuration on this loop thus defines a map from the physical-space $S^1$ to the vacuum-manifold $S^1$, characterized by an integer winding number via $\pi_1(S^1)=\mathbb{Z}$. A nonzero winding number indicates a topologically nontrivial phase configuration corresponding to a vortex, while a winding number of zero corresponds to a trivial configuration with no vortex.

\begin{figure}[!htbp]
    \centering
    \includegraphics[width=0.35\textwidth]{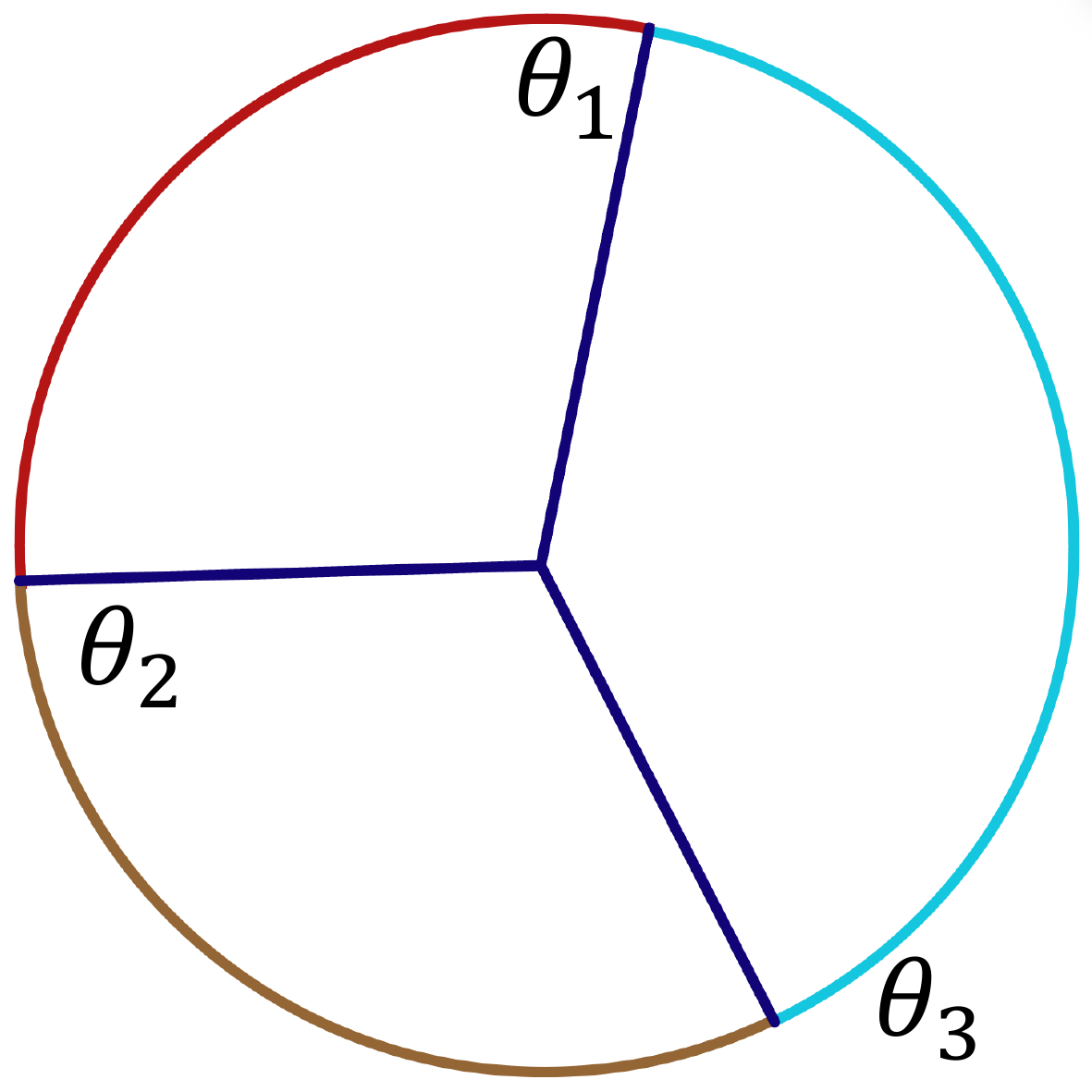} 
    \hspace{1.5cm}
    \includegraphics[width=0.35\textwidth]{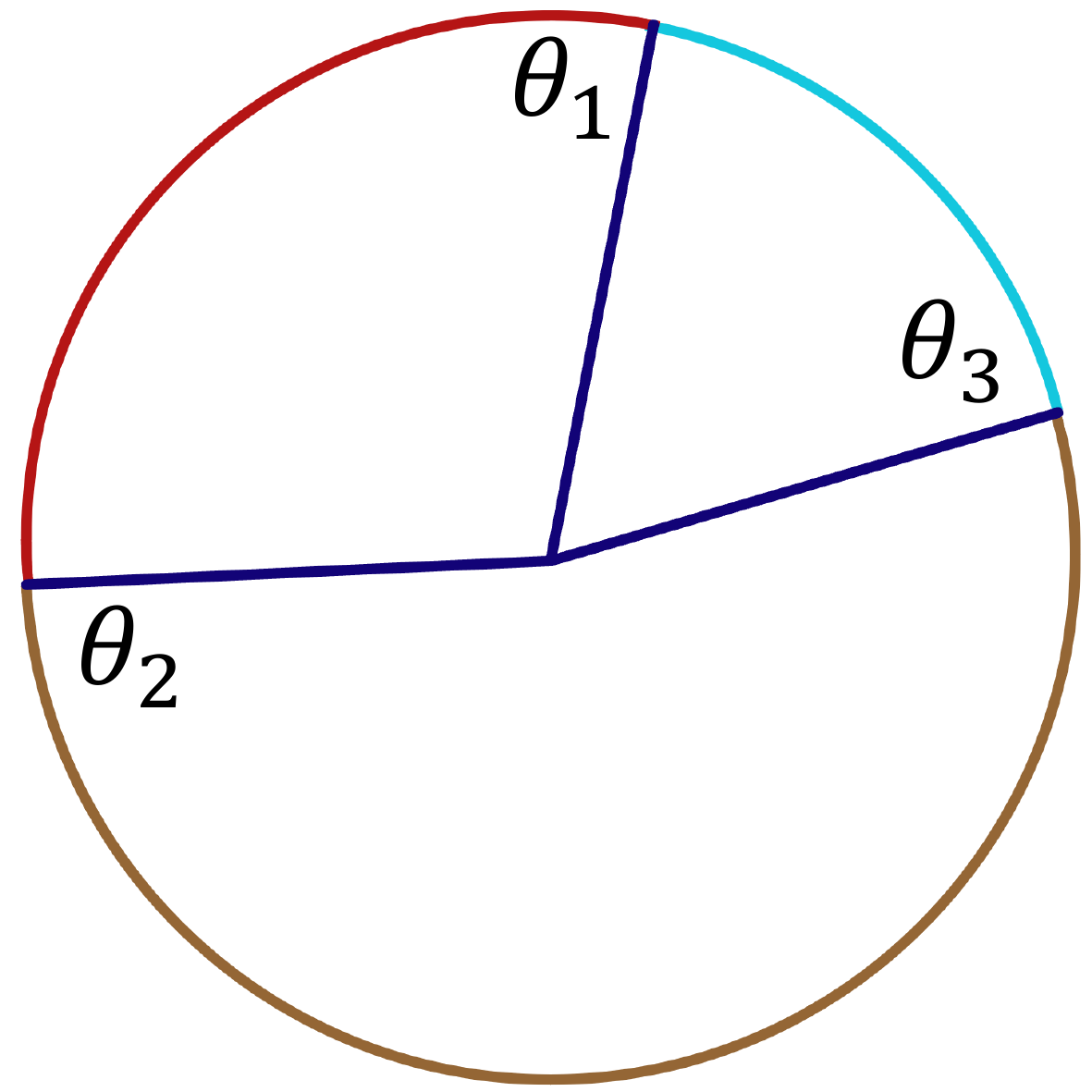}

    \caption{Schematic of the geodesic rule applied to phase interpolation during three-bubble collision. The vacuum manifold of the $U(1)$ symmetry-broken phase is a circle $S^1$, with bubble phases $\theta_1,\theta_2,\theta_3$ as points on it. \textbf{Left}: When all three arcs between adjacent phases are shorter than $\pi$, the geodesic rule chooses the shorter arc for phase interpolation, resulting in a net winding number and vortex formation. \textbf{Right}: When one arc (brown) exceeds $\pi$, that arc is not used; the interpolation path does not encircle the full manifold, yielding zero winding and no vortex.}  
    \label{geo_explain}
\end{figure}

To illustrate the vortex formation process, we simulate the collision of three bubbles with phases $\theta_1$, $\theta_2$, and $\theta_3$. For a transition that spontaneously breaks a $U(1)$ symmetry, the vacuum manifold is a circle $S^1$, and each bubble's phase corresponds to a point on this circle, as shown in Fig.~\ref{geo_explain}. These three points naturally divide the circle into three arcs. According to the geodesic rule, the phase difference between any two bubbles is interpolated along the shorter arc at their junction. Consequently, a non-zero net winding number arises when the three points are distributed such that all three connecting arcs are shorter than $\pi$. In this configuration, the phase interpolation around the junction of the three bubbles traces a closed loop covering the entire vacuum manifold $S^1$, as illustrated in the left panel of Fig.~\ref{geo_explain}. Conversely, consider the case where one of the arcs defined by $\theta_1$, $\theta_2$, and $\theta_3$ exceeds $\pi$, as highlighted by the brown arc in the right panel of Fig.~\ref{geo_explain}. According to the geodesic rule, such a large arc represents an energetically unfavorable interpolation path and is therefore not selected during phase relaxation when the bubbles coalesce. As a result, no net winding number is generated in this configuration.


\begin{figure}[!htbp]
    \centering
    \includegraphics[width=0.9\textwidth]{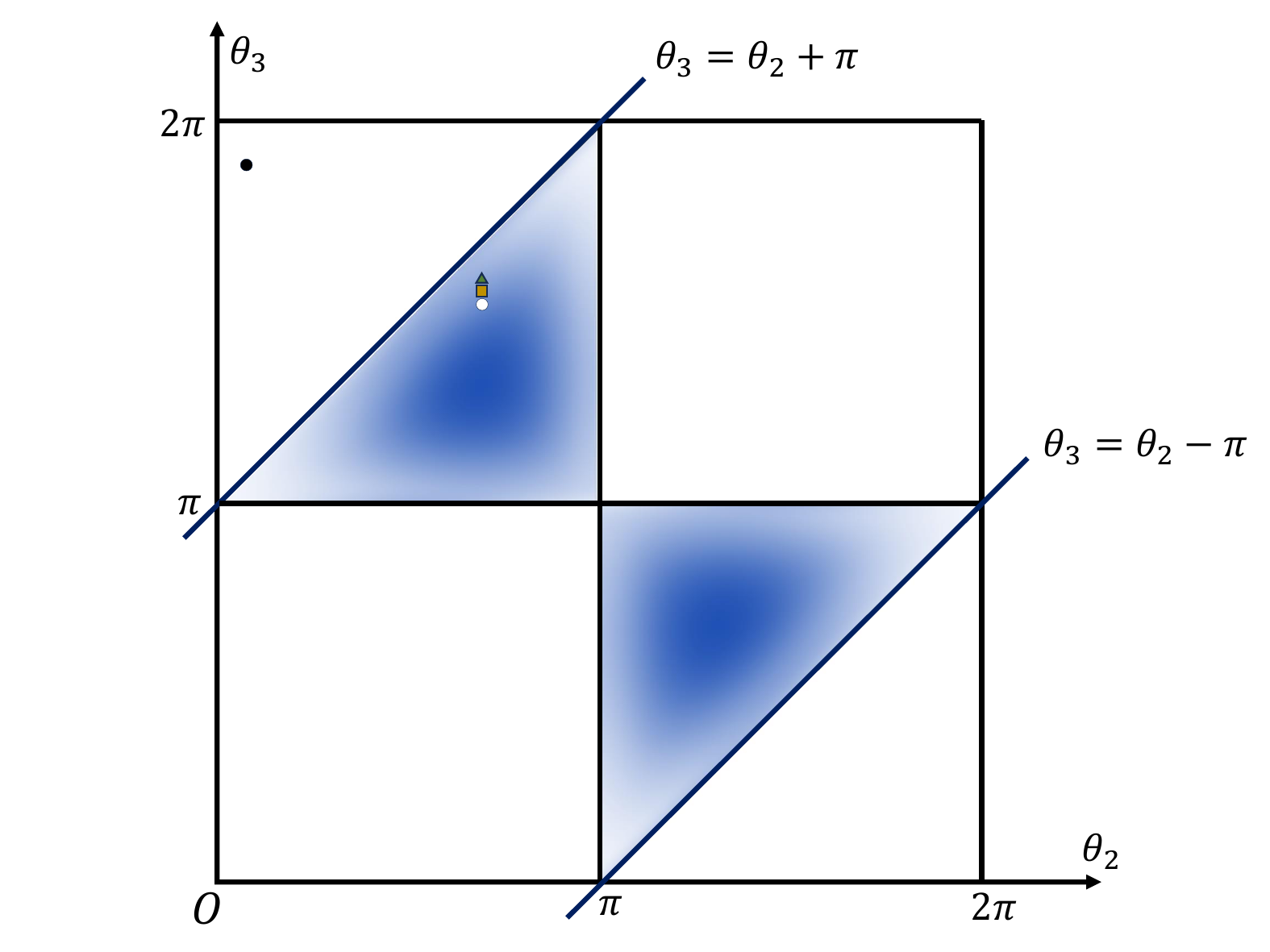}
    \caption{Parameter space for vortex formation with $\theta_1$ fixed at $0$. According to the geodesic rule, a vortex forms if the point $(\theta_2,\theta_3)$ lies within the blue region. The final state is highly sensitive to the initial phases. The color gradient indicates that initial conditions near the center of the triangular region (darker blue) have a higher probability of yielding a single vortex, whereas those near the boundary typically result in a vortex-free state. Markers denote the initial phase differences of the time-evolving examples shown in subsequent figures.}  
    \label{geo_probability}
\end{figure}

The above analysis shows that a vortex is expected to form when the three phases partition the vacuum manifold $S^1$ into three arcs, each shorter than $\pi$. Geometrically, this condition is equivalent to the requirement that the three phase points on $S^1$ are not all contained within any semicircle. Without loss of generality, we set $\theta_1 = 0$ by a global phase rotation. The condition for vortex formation can then be stated as follows: if $\theta_2 > \pi$, then $\theta_3$ must lie in the interval $(\theta_2 - \pi, \pi)$; if $0 < \theta_2 < \pi$, then $\theta_3$ must lie in $(\pi, \theta_2 + \pi)$. The region in the $\theta_2$-$\theta_3$ plane corresponding to vortex formation is shaded in blue in Fig.~\ref{geo_probability}. Assuming the phases of the three bubbles are random and uncorrelated, the probability of forming a vortex  is therefore $1/4$.


\begin{figure}[!htbp]
    \centering
    \includegraphics[width=0.99\textwidth]{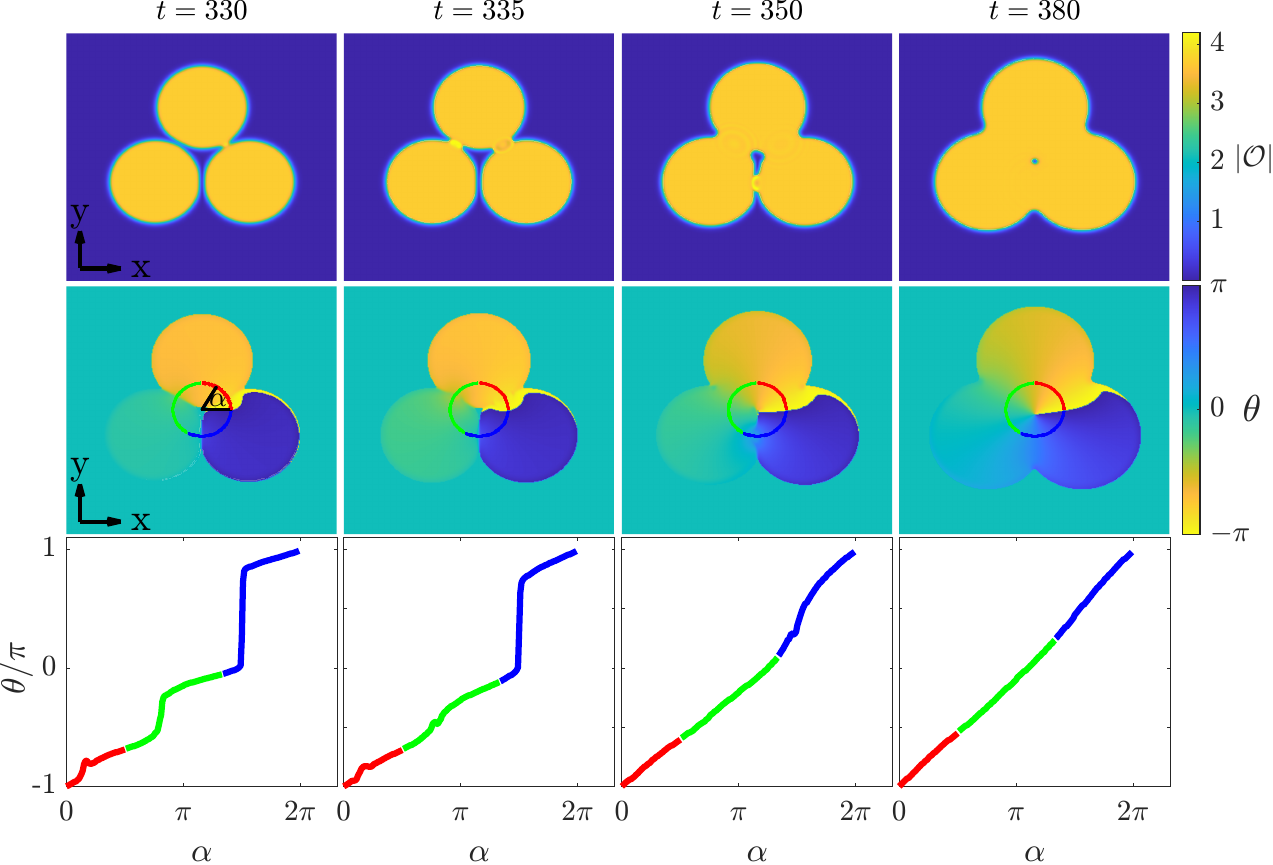}
    \caption{Condensate amplitude and phase distributions at representative times. Initial bubble phases: $\theta_1 = 0$, $\theta_2 = 2\pi/3$, $\theta_3 = (4+0.7)\pi/3$ (orange square in Fig.~\ref{geo_probability}). \textbf{Top row}: condensate amplitude $|\mathcal{O}|$. \textbf{Middle row}: corresponding phase $\arg(\mathcal{O})$. \textbf{Bottom row}: phase profiles extracted from the colored ring in the middle row, with colors matching the arcs. The spatial region is $[0,150]\times[0,150]$ with the parameters $\rho = 3.7$, $\lambda = -2$, $\tau = 0.8$, $\sigma = 5$, $h = 1.25$.}  
    \label{geo_defect}
\end{figure}

To investigate the validity of the geodesic rule under non-equilibrium conditions, we numerically simulate the collision of three bubbles nucleated in the normal phase. The initial phases are chosen as $\theta_1 = 0$, $\theta_2 = 2\pi/3$, $\theta_3 = (4 + 0.7)\pi/3$ (orange square in Fig.~\ref{geo_probability}), a configuration that the geodesic rule predicts will generate a vortex. The real-time dynamics of the bubble collision are shown in Fig.~\ref{geo_defect}. The top row displays the condensate distribution, and the middle row shows the corresponding phase distribution. To track the evolution of the phase structure during the collision, we extract phase values along colored circles superimposed on the phase field. These circles are chosen to pass through the initial collision region—the area where the bubble boundaries first meet. Denoting the angular coordinate along each circle by $\alpha$, we plot the extracted phase profiles in the bottom row of Fig.~\ref{geo_defect}. Along the colored ring, the phases of the three bubbles exhibit smooth interpolation across their junctions. Crucially, this interpolation follows the minor arcs of the vacuum manifold $S^1$. The bubbles are nucleated symmetrically, specifically positioned at the vertices of an equilateral triangle, so that the collision point between any two bubbles lies at the midpoint of the line connecting their nucleation centers. Interestingly, as the bubbles approach and their boundaries interact, the sequence of collisions depends on the initial phase differences, thereby breaking the symmetry. More precisely, we find that the pair of bubbles with the smallest phase difference (traced by the red segment of the circle) collides first at \(t = 330\), followed by the pair with the next smallest difference (green segment) at \(t = 335\). An asymmetric collision could produce a nonzero momentum for the vortex.~\footnote{This feature is more visible in the first snapshot (at $t=215$) of Fig.~\ref{geo_condensate_1d} below.} Upon collision, the phase difference within each newly merged domain smooths out rapidly, as illustrated by the evolution of the red and green segments of the phase profiles in the third row of Fig.~\ref{geo_defect}. By \(t = 380\), a vortex has already formed near the center of the three bubbles, consistent with the prediction of the geodesic rule. Identifying vortices in a dynamically evolving system is not a trivial problem. The method we use to identify vortices from numerical simulations is described in Appendix~\ref{app:B}, which combines the phase winding analysis with the temporal behavior of the condensate magnitude.

\begin{figure}[!htbp]
    \centering
    \includegraphics[width=1\textwidth]{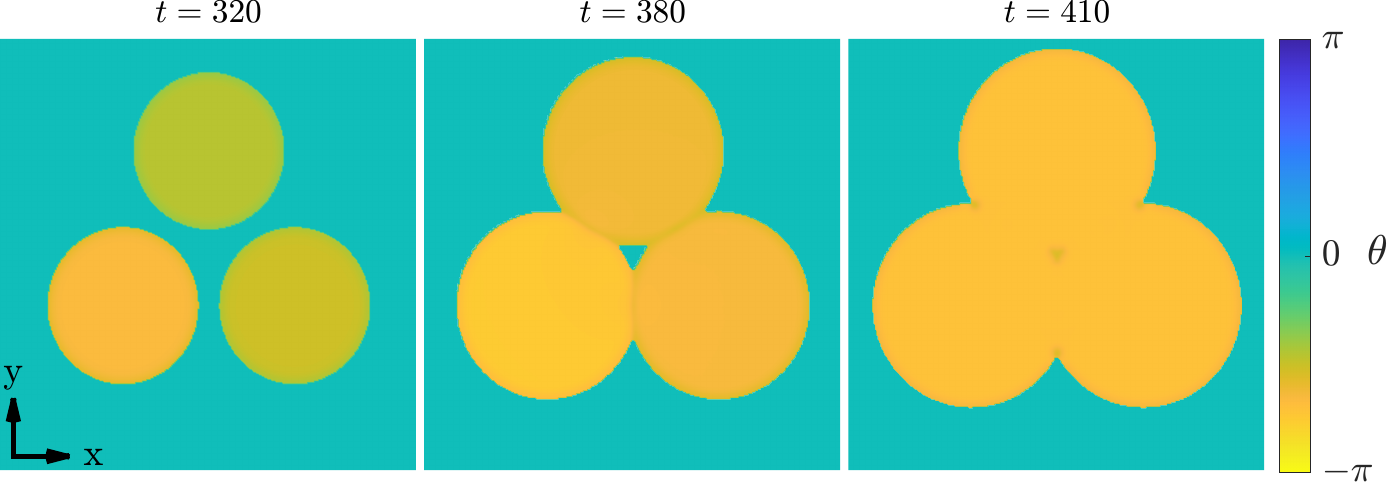}
   
    \caption{Phase configurations at three representative times. Initial bubble phases are set to $0.1\pi$, $1.9\pi$, and $0.05\pi$ (black circle in Fig.~\ref{geo_probability}; shifted by $-0.05\pi$ to align with $\theta_1=0$), following the same configuration as in~\cite{Digal:1996ip} where vortex-antivortex pair production was observed. In contrast, our holographic system exhibits no such pair production throughout the entire evolution. We have considered parameters $\rho = 3.7$, $\lambda = -2$, $\tau = 0.8$, $\sigma = 5$, $h = 1.25$}
    \label{fig:geo_compare_phase}
\end{figure}

Before proceeding, we note that a breakdown of the geodesic rule was reported in~\cite{Digal:1996ip} for a $U(1)$ global theory in the absence of any explicit symmetry breaking in 2+1 dimensions. It was shown that sufficiently energetic oscillations of the complex scalar field during bubble collisions can drive the field over the potential barrier, leading to vortex–antivortex pair production. In particular, the study of~\cite{Digal:1996ip} considered pair generation for initial phase differences of $0.1\pi$, $1.9\pi$, and $0.05\pi$. To set one of the phases to zero (consistent with the parameter space in Fig.~\ref{geo_probability}, where $\theta_1 = 0$), we subtract $0.05\pi$ from each value. This configuration then corresponds to the black circle in Fig.~\ref{geo_probability}, for which the geodesic rule would predict no topological vortex formation. In contrast to the high-speed collisions examined in~\cite{Digal:1996ip}, the bubble expansion velocity in our setup is relatively slow due to strong coupling between the expanding bubble wall and the surrounding thermal plasma (see Fig.~\ref{rho_v_scale}). As a result, field oscillations remain weak, and defect-antidefect pair production via field oscillations does not occur in our holographic model (see Fig.~\ref{fig:geo_compare_phase}).

Somewhat unexpectedly, vortex-antivortex pair production does occur in our system for initial phase configurations that lie within the blue region of Fig.~\ref{geo_probability} (obtained from the geodesic rule). To demonstrate this, we simulate a representative case with initial phases $\theta_1 = 0$, $\theta_2 = 2\pi/3$, $\theta_3 = (4 + 0.8)\pi/3$ (indicated by the green triangle in Fig.~\ref{geo_probability}). The resulting collision dynamics are displayed in Fig.~\ref{geo_condensate_1d}. The first and second rows show the condensate distribution and the corresponding phase field, respectively. By $t = 215$, a vortex has already formed, consistent with the geodesic rule. Notably, however, the first vortex does not emerge at the center of the three bubbles; instead, it appears closer to the boundary and acquires a nonzero velocity. More strikingly, an antivortex emerges near the boundary of the bubble carrying the largest initial phase difference, giving rise to a vortex–antivortex pair, as clearly seen in the snapshot at $t = 275$. As time evolves, the vortex and antivortex gradually approach each other and eventually annihilate, leaving a final state devoid of any vortex. This process thus reduces the probability of reaching a stable state containing a single vortex, beyond the naive expectation based solely on the geodesic rule.

\begin{figure}[!htbp]
    \centering
    \includegraphics[width=0.96\textwidth]{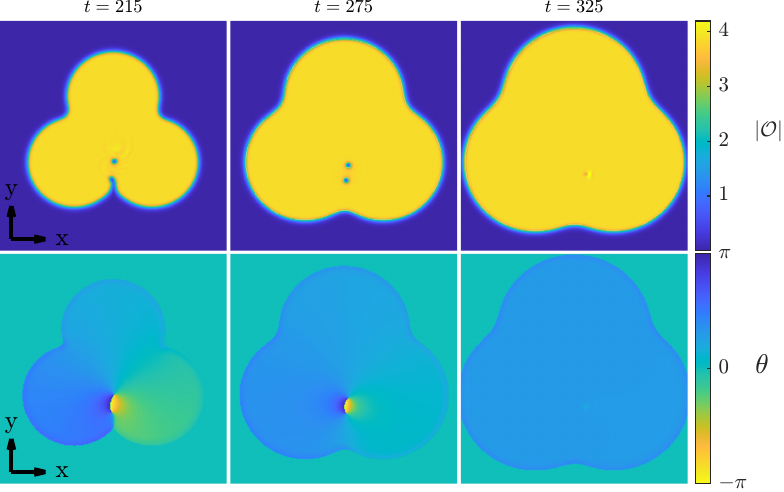}
    \caption{Condensate amplitude and phase distributions at three representative times. Initial bubble phases: $\theta_1 = 0$, $\theta_2 = 2\pi/3$, $\theta_3 = (4+0.8)\pi/3$ (green triangle in Fig.~\ref{geo_probability}). \textbf{Top row}: condensate amplitude $|\mathcal{O}|$ at $t = 215$, $275$, and $380$ (left to right). \textbf{Bottom row}: corresponding phase $\arg(\mathcal{O})$. The plotted spatial region is $[40,160]\times[40,160]$. We have set $\rho = 3.8$, $\lambda = -2$, $\tau = 0.8$, $\sigma = 5$, $h = 1.25$.}  
    \label{geo_condensate_1d}
\end{figure}

\begin{figure}[htbp]
    \centering        \includegraphics[width=0.6\textwidth]{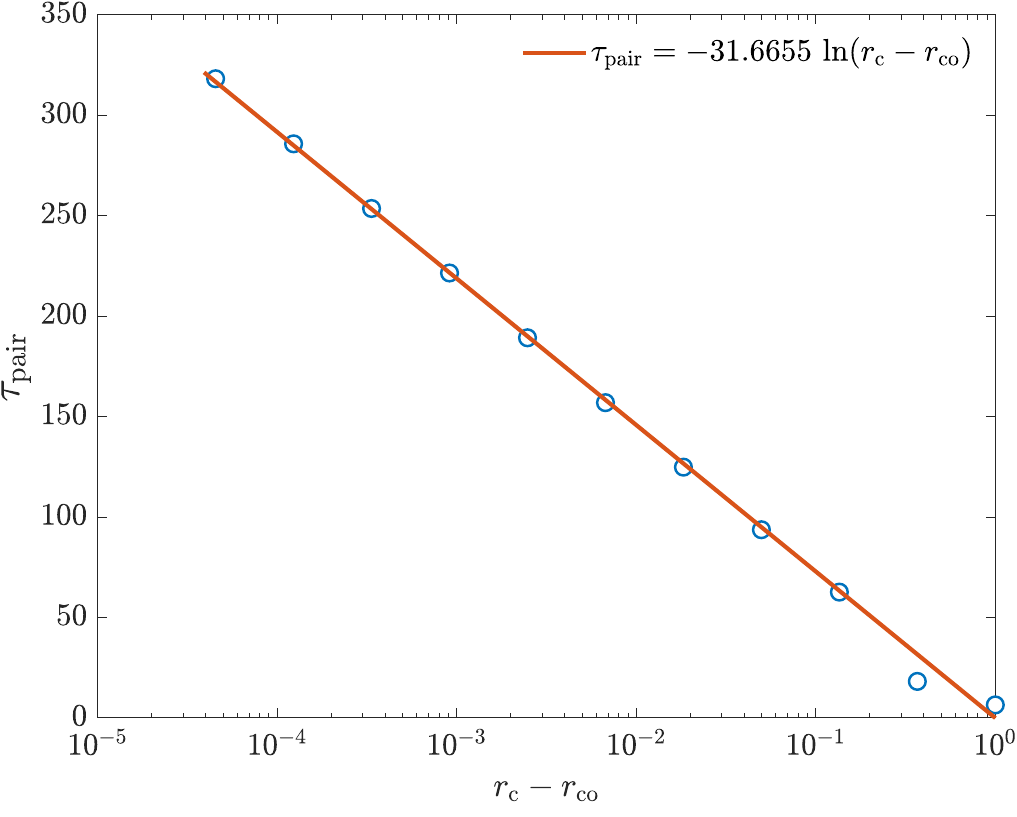}   
    \caption{Lifetime of the vortex-antivortex pair, $\tau_{\mathrm{pair}}$, as a function of the three-bubble collision radius $r_{\mathrm{co}}$, with parameters fixed at $\rho = 3.8$, $\lambda = -2.08$, $\tau = 0.8$, $\sigma = 5$, $h = 1.25$. The critical radius is $r_c \approx 12.9825$. Blue circles represent the numerical data, and the red line is the corresponding fit. The lifetime exhibits a linear dependence on $\ln(r_c - r_{\mathrm{co}})$, given by $\tau_{\mathrm{pair}} = -31.6655 \ln(r_{\mathrm c} - r_{\mathrm{co}})$.}
    \label{doublev_tau_r_scale}
\end{figure}

A key question is to what extent these phenomena exist in the three bubble collision process. To quantify this, we systematically vary the bubble collision velocity $v_{\text{co}}$ and collision radius $r_{\text{co}}$ by adjusting both the nucleation positions and the coupling parameter $\lambda$. The initial phases are set to $\theta_1 = 0$, $\theta_2 = 2\pi/3$, $\theta_3 = (4 + 0.8)\pi/3$ (green triangle in Fig.~\ref{geo_probability}). The bubbles are nucleated symmetrically, with their centers positioned at the vertices of an equilateral triangle; consequently, each pair of bubbles collides at the midpoint of the line connecting their nucleation centers. We define $r_{\text{co}}$ as half of the distance between two nucleation centers. The bubble-wall velocity at collision is obtained from single-bubble expansion data, as the radius and wall velocity are in one-to-one correspondence once all other parameters are fixed. To investigate this process quantitatively, we vary the collision radius $r_{\mathrm{co}}$ while keeping the coupling parameter fixed at $\lambda=-2.08$. By means of a bisection search, we identify a critical radius, $r_c \approx 12.9825$, such that the system evolves to a vortex-free terminal state for $r_{\mathrm{co}}<r_c$, while for $r_{\mathrm{co}}>r_c$ it evolves to a terminal state containing a single vortex. We find that vortex-antivortex pair production occurs most prominently when the initial collision radius is slightly below the critical value. In addition, Fig.~\ref{doublev_tau_r_scale} shows the lifetime of the vortex-antivortex pair, denoted by $\tau_{\mathrm{pair}}$, as a fucnction of $r_c-r_{\mathrm{co}}$. Near the critical radius, $\tau_{\mathrm{pair}}$ increases linearly with $\ln (r_{\mathrm c}-r_{\mathrm{co}})$.~\footnote{Due to limitations of computational power, we do not extend the calculation to smaller values of $r_c-r_{\mathrm{co}}$, because a larger $\tau_{\mathrm{pair}}$ implies a longer bubble-expansion time, which in turn requires a larger computational domain and substantially greater numerical resources.}

\begin{figure}[!htbp]
    \centering
    \includegraphics[width=0.6\textwidth]{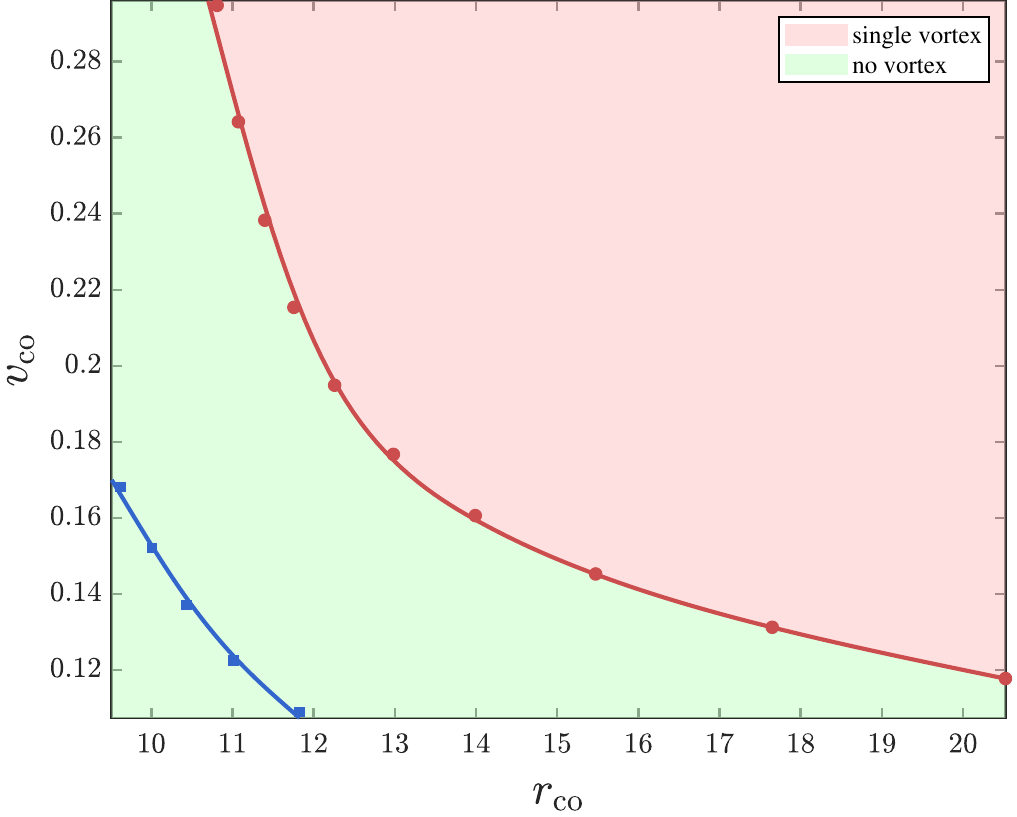}
    \caption{Phase diagram for vortex generation in the $(v_{\mathrm{co}},r_{\mathrm{co}})$ parameter space, with $\rho = 3.8$, $\tau = 0.8$, $\sigma = 5$, and $\lambda$ varied to access a range of bubble-wall velocities. Two initial phase configurations are considered: (I) $\theta_3 = (4+0.8)\pi/3$ and (II) $\theta_3 = (4+0.6)\pi/3$, with $\theta_1 = 0$, $\theta_2 = 2\pi/3$ fixed (green triangle and white circle in Fig.~\ref{geo_probability}, respectively). Red circles and blue squares denote critical values for cases I and II; solid red and blue lines are spline fits. The red region (above the red curve) yields a single stable vortex for case I, while the green region below yields no vortex. Below the red curve, adjacent to the critical boundary, a transitional zone hosts long-lived vortex-antivortex pairs for case I (see Fig.~\ref{doublev_tau_r_scale}). The blue curve shifts downward and leftward compared with the red curve, expanding the single-vortex region for case II. This indicates that initial phases located near the center of the triangular region in Fig.~\ref{geo_probability} are more favorable for single-vortex formation than those near the edges. }  
    \label{geodestroy_rv}
\end{figure}

We further investigate how the bubble‑wall velocity influences vortex generation under the phase configuration $\theta_1 = 0$, $\theta_2 = 2\pi/3$, $\theta_3 = (4 + 0.8)\pi/3$. To this end, we vary the parameter $\lambda$, since different values of $\lambda$ lead to different relationships between the bubble‑wall velocity and the bubble radius, thereby providing a wide parameter range for our calculations. The resulting phase diagram in the $(v_{\text{co}}, r_{\text{co}})$ plane is shown in Fig.~\ref{geodestroy_rv}. The red point marks the critical values obtained at a specific $\lambda$. Using spline interpolation on these points, we obtain a critical boundary (solid red line in Fig.~\ref{geodestroy_rv}). The red region corresponds to parameter sets that yield a final state containing a single stable vortex; its lower edge is delineated by the solid red line. Below this boundary, in the green region, the terminal state contains no vortex. Adjacent to the critical line lies a transitional zone where long‑lived vortex–antivortex pairs appear (see the lifetime measurements in Fig.~\ref{doublev_tau_r_scale}). We find that both a high collision velocity and a large collision radius significantly enhance the probability of evolving into a single‑vortex state.

To examine the sensitivity to initial phases, we compute the critical boundary (solid blue line in Fig.~\ref{geodestroy_rv}) for a slightly modified phase configuration, corresponding to the white circle in Fig.~\ref{geo_probability}: $\theta_1 = 0$, $\theta_2 = 2\pi/3$, $\theta_3 = (4+0.6)\pi/3$. Compared with the previous case ($\theta_3 = (4+0.8)\pi/3$), this adjustment shifts $\theta_3$ closer to the center of the triangular region in Fig.~\ref{geo_probability}. The blue squares mark the critical values obtained at different $\lambda$ values, and spline interpolation again yields the critical line. Remarkably, the resulting critical line moves downward and leftward, thereby expanding the parameter region that leads to a single‑vortex final state. This shift clearly demonstrates that initial phases located near the center of the triangle are more favorable for single‑vortex formation than those near the edges. Consequently, in Fig.~\ref{geo_probability} we employ a color gradient to represent this probability: not all phase configurations within the shaded region guarantee a single‑vortex outcome. Instead, the likelihood of reaching a stable single‑vortex state varies continuously. Parameters near the center exhibit a higher probability, whereas those near the boundary more often result in vortex‑free final states.

\section{Conclusion}
\label{conclusion}

In this work, we have systematically investigated bubble dynamics in a holographic superfluid exhibiting a first‑order phase transition with spontaneous $U(1)$ symmetry breaking, with particular emphasis on vortex formation driven by multi‑bubble collisions. We first identified universal critical behavior near the bubble nucleation threshold. By fine‑tuning the initial perturbation amplitude $h$ to the critical value $h_c$, the system exhibits a prolonged near‑critical stage, common to both supercritical and subcritical evolutions. The departure from the critical solution is governed by a single unstable mode, leading to a logarithmic scaling law $\tau_{\text{scale}} \sim -\ln |h - h_c|$ (right panel of Fig.~\ref{fig:total}). This behavior mirrors critical phenomena in gravitational collapse and black hole formation, reinforcing the deep analogy between holographic phase transitions and gravitational criticality. Beyond the critical regime, we quantitatively studied the bubble wall velocity. For a fixed temperature, the terminal velocity increases with the charge density $\rho$ but remains relatively small, reflecting the strong dissipative coupling between the expanding bubble wall and the surrounding plasma. 

Turning to multi‑bubble collisions, we examined vortex formation under the geodesic rule. While the rule predicts a $1/4$ probability for a single vortex when three bubbles merge with random phases, our simulations reveal significant deviations under non‑equilibrium conditions. In particular, for initial phases satisfying the vortex‑formation condition, we observed the emergence of a vortex‑antivortex pair, which subsequently annihilates and leaves a vortex‑free final state. This process reduces the likelihood of a stable single vortex beyond the geodesic prediction. Such pair production occurs predominantly when the collision radius is slightly below a critical value $r_c$ and is accompanied by a linear scaling of the pair lifetime with $\ln(r_{\mathrm c} - r_{\text{co}})$ (see Fig.~\ref{doublev_tau_r_scale}).

We further mapped out the phase diagram for vortex formation in the $(v_{\text{co}}, r_{\text{co}})$ plane (see Fig.~\ref{geodestroy_rv}). A critical boundary separates the region that yields a single stable vortex from the vortex‑free region. Adjacent to this boundary, a transitional zone hosts long‑lived vortex‑antivortex pairs. Both higher collision velocity and larger collision radius significantly enhance the probability of reaching a single‑vortex final state, highlighting the importance of dynamical parameters beyond the geodesic rule. The sensitivity to initial phases was examined. It was found that the vortex formation probability varies continuously within the shaded region of Fig.~\ref{geo_probability}: phase configurations near the center favor a single vortex, while those near the edge more often lead to vortex‑free outcomes. Consequently, the color gradient in Fig.~\ref{geo_probability} qualitatively captures this varying likelihood. 

Our holographic study reveals that bubble dynamics and vortex formation in a first‑order superfluid phase transition are governed by rich non‑equilibrium phenomena. In particular, the newly observed process of vortex–antivortex pair production and annihilation makes it more challenging to understand the KZM in the context of first-order phase transitions. These results underscore the necessity of full dynamical simulations in strongly coupled systems and provide a quantitative foundation for understanding topological defect formation in first‑order phase transitions, with potential implications for early‑universe cosmology and quantum matter. Several directions warrant further investigation. First, we have shown that the probability of reaching a terminal state containing a single vortex depends on the initial phase differences. A systematic study of this probability distribution would be valuable, particularly to determine whether it may have observable implications in other contexts, such as cosmology. Second, the mechanism responsible for vortex–antivortex pair production in our setup remains unclear. It would be worthwhile to investigate whether a similar phenomenon can also occur during domain merging in second-order phase transitions. A related open question concerns the origin of the scaling behavior exhibited by the lifetime of the vortex–antivortex pair near the critical radius $r_c$. Third, the setup considered here is relatively special, since the three bubbles are nucleated symmetrically. It is therefore natural to ask whether similar phenomena persist when the bubbles are nucleated in a less symmetric configuration. One may also investigate whether vortices with higher winding numbers can be generated when more bubbles participate in the collision. More broadly, it would be interesting to explore systems with different symmetry-breaking patterns, such as p-wave system~\cite{Cai:2013pda,Cai:2013aca}. We leave these questions to future work.

\section{Acknowledgements}
This work is supported by the National Natural Science Foundation of China Grants No.\,12525503, No.\,12588101 and No.\,12447101. We acknowledge the use of the High Performance Cluster at the Institute of Theoretical Physics, Chinese Academy of Sciences.

\appendix
\section{Numerical scheme of the fully non-linear simulations}\label{app:A}

We solve the full non-linear equations of motion for the matter fields in the probe limit, where the background geometry is fixed to the AdS-Schwarzschild black brane~\eqref{AdSsw}. The equations of motion derived from the action are
\begin{align}
D_\mu D^\mu \Psi &= m^2 \Psi + 2\lambda (\Psi^*\Psi)\Psi + 3\tau (\Psi^*\Psi)^2 \Psi\,, \\
\nabla_\mu F^{\mu\nu} &= -2\,\text{Im}\left(\Psi^* D^\nu \Psi\right),
\end{align}
where $\text{Im}$ denotes the imaginary part. To facilitate numerical integration, we introduce the rescaled scalar field $\Phi \equiv \Psi / z$. The equations can be written explicitly as a set of PDEs of first order in time. After straightforward but lengthy algebra, we obtain:
\begin{equation}
\begin{split}
    2\partial_t\partial_z\Phi-[2iA_t\partial_z\Phi+i\partial_zA_t\Phi+\partial_z(f\partial_z\Phi)-z\Phi+\partial_x^2\Phi+\partial_y^2\Phi-i(\partial_xA_x+\partial_yA_y)\Phi\\
    -(A_x^2+A_y^2)\Phi-2i(A_x\partial_x\Phi+A_y\partial_y\Phi)-2\lambda\Phi^*\Phi\Phi-3z^2\tau(\Phi^*\Phi)^2\Phi]=0\,,\label{phi}
\end{split}
\end{equation}

\begin{equation}
\begin{split}
    2\partial_t\partial_zA_x-[\partial_z(\partial_xA_t+f\partial_zA_x)+\partial_y(\partial_yA_x-\partial_xA_y)-2A_x|\Phi|^2+2\text{Im}(\Phi^*\partial_x\Phi)]=0\,,\label{Ax}
\end{split}
\end{equation}

\begin{equation}
\begin{split}
    2\partial_t\partial_zA_y-[\partial_z(\partial_yA_t+f\partial_zA_y)+\partial_x(\partial_xA_y-\partial_yA_x)-2A_y|\Phi|^2+2\mathrm{Im}(\Phi^*\partial_y\Phi)]=0\,,\label{Ay}
\end{split}
\end{equation}

\begin{equation}
\begin{split}
    \begin{aligned}\partial_t\partial_zA_t-[\partial_x^2A_t+\partial_y^2A_t+f\partial_z(\partial_xA_x+\partial_yA_y)-\partial_t(\partial_xA_x+\partial_yA_y)-2A_t|\Phi|^2\\-2f\text{Im}(\Phi^*\partial_z\Phi)+2\text{Im}(\Phi^*\partial_t\Phi)]=0\,,\end{aligned}\label{At}
\end{split}
\end{equation}

\begin{equation}
\begin{split}
    \partial_z(\partial_xA_x+\partial_yA_y-\partial_zA_t)-2\mathrm{Im}(\Phi^*\partial_z\Phi)=0\,.
    \label{constraint}
\end{split}
\end{equation}
The dynamical variables are $\{\Phi, A_{x}, A_{y}, A_{t}\}$. The last one~\eqref{constraint} is a constraint that must be satisfied at all times; it is used to determine $A_t$ once the other fields are known.

We impose periodic boundary conditions in the spatial directions $x$ and $y$, and discretize these directions using the Fourier pseudo-spectral method. In the radial direction $z$, we employ the Chebyshev pseudo-spectral method with grid points clustered near the AdS boundary. At the AdS boundary $z=0$, we impose for all times:
\begin{align}
A_x|_{z=0} = A_y|_{z=0} = 0, \qquad \partial_z A_t|_{z=0} = -\rho, \qquad \Phi|_{z=0} = 0,
\end{align}
where $\rho$ is the fixed charge density of the boundary system. The condition $\Phi|_{z=0}=0$ enforces spontaneous symmetry breaking (no source for the dual operator). At the black hole horizon $z=z_h$, no additional boundary condition is required because the ingoing Eddington-Finkelstein coordinates ensure regularity; the evolution equations are simply integrated through the horizon.

To trigger bubble nucleation, we introduce a small perturbation to the condensate at the initial time $t=0$. This is implemented by modifying the near-boundary behavior of $\Phi$:
\begin{equation}
\Phi(t=0) \rightarrow \Phi(t=0) + z\,\delta\mathcal{O}(x,y)\,,
\end{equation}
where $\delta\mathcal{O}(x,y)$ is a Gaussian wave packet centered at the desired nucleation position(s). For single-bubble simulations we use the profile given in~\eqref{gaussian_wave}; for multi-bubble cases we superpose several such packets.

The time evolution proceeds by solving the dynamical equations~\eqref{phi}--\eqref{Ay} for $\Phi$, $A_x$, $A_y$ using a fourth-order Runge-Kutta scheme. At each time step, we then determine $A_t$ by solving the constraint equation \eqref{constraint}. This requires knowledge of $A_t|_{z=0}$ and $\partial_z A_t|_{z=0}$. The latter is fixed by the boundary condition $\partial_z A_t|_{z=0} = -\rho$. The value of $A_t$ at the boundary, which is related to the chemical potential in the dual field theory, is obtained by evaluating~\eqref{At} at $z=0$. After using the boundary conditions, we find
\begin{equation}
\left[\partial_x^2 A_t + \partial_y^2 A_t + f\partial_z(\partial_x A_x + \partial_y A_y)\right]_{z=0} = 0\,.
\end{equation}
This elliptic equation is solved in Fourier space. The zero mode of $A_t|_{z=0}$ is set to zero, which corresponds to a choice of gauge (\emph{i.e.}, we fix the spatial average of $A_t|_{z=0}$ to zero).

The simulation domain sizes are chosen depending on the physical process under study. For the critical behavior analysis (Section~\ref{nucleation and critical}), we use $L_x \times L_y = 80 \times 80$ discretized with $241 \times 241$ grid points. For the bubble wall velocity study (Section~\ref{bubble_wall_v}), a larger domain $L_x \times L_y = 330 \times 330$ with $831 \times 831$ grid points is used to accommodate the expanding bubble. For the vortex formation analysis (Section~\ref{defect_generation}), the domain is $L_x \times L_y = 200 \times 200$ with $401 \times 401$ grid points. In all cases, the domain size is sufficiently large so that boundary effects do not influence the physical phenomena of interest. The radial direction $z$ is discretized with $N_z = 20$ grid points, which provides adequate resolution for the near-boundary dynamics. Time steps are typically $\Delta t = 0.01$ for the critical and vortex studies, and $\Delta t = 0.05$ for the wall velocity runs, ensuring numerical stability and convergence.

\section{Vortex identification}\label{app:B}

Identifying vortices in a dynamically evolving superfluid condensate requires careful analysis, as both the phase winding and the condensate magnitude must be considered. A vortex is characterized by a nontrivial winding of the phase of the order parameter around a point, accompanied by a vanishing condensate amplitude at the core. In this work, we adopt two complementary criteria to reliably detect vortices from our numerical simulations.

\begin{figure}[!htbp]
    \centering
    \includegraphics[width=0.4\textwidth]{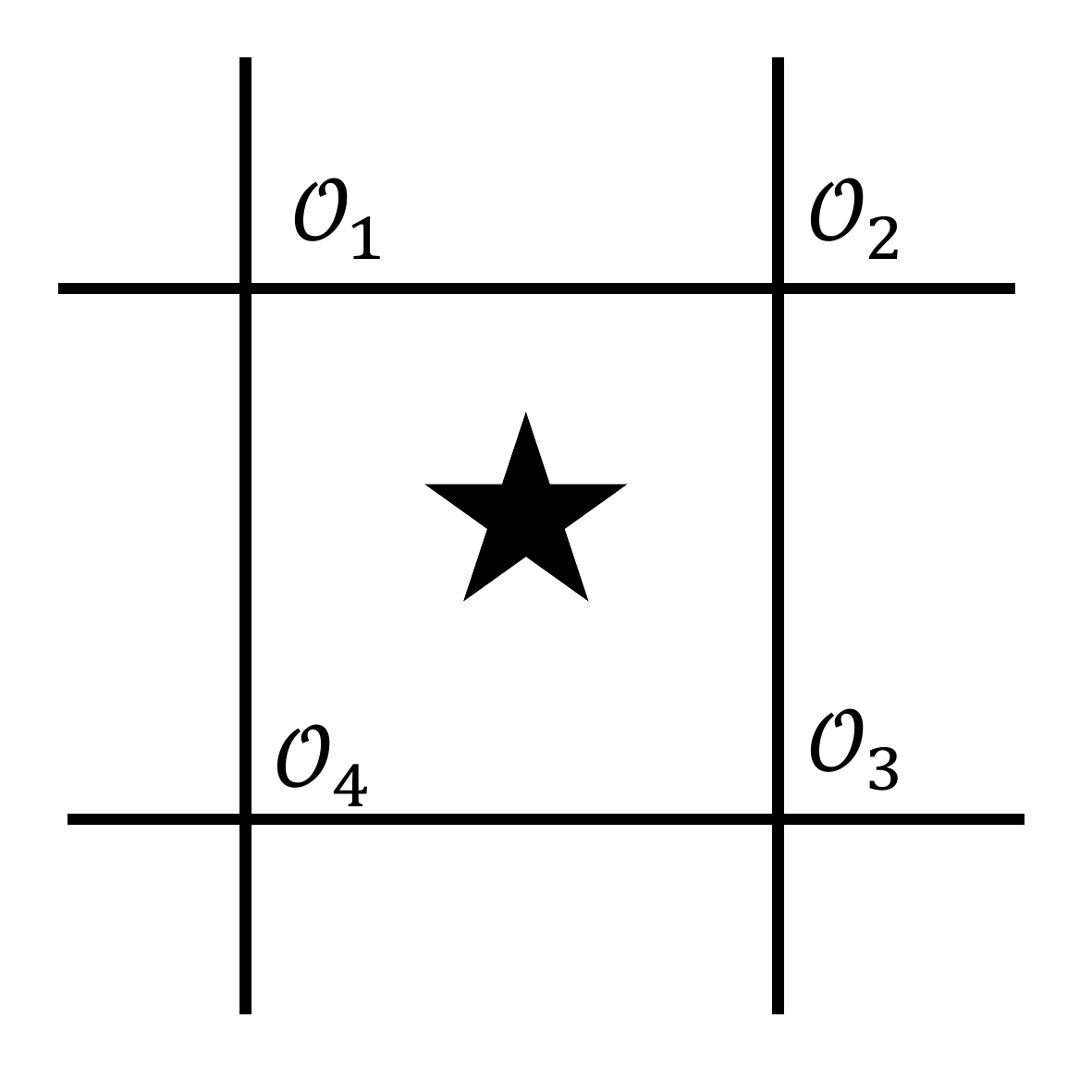}
    \caption{Illustration of the vortex-identification procedure on the grid. We evaluate the condensate values $\mathcal{O}_1$, $\mathcal{O}_2$, $\mathcal{O}_3$, and $\mathcal{O}_4$ at four corners of a grid square. The winding number is computed from the phase differences around each plaquette.}
    \label{vortex_identification}
\end{figure}

The first criterion is based on the topological winding number. For a spontaneously broken $U(1)$ symmetry, the phase of the condensate $\arg(\mathcal{O})$ must wind by an integer multiple of $2\pi$ around a vortex core. To compute the winding number numerically, we examine each elementary plaquette formed by four neighboring grid points, as illustrated in Fig.~\ref{vortex_identification}. Let $\mathcal{O}_1$, $\mathcal{O}_2$, $\mathcal{O}_3$, $\mathcal{O}_4$ be the condensate values at the four corners of a plaquette, ordered counterclockwise. The total phase change around the plaquette is defined as
\begin{equation}
    \Delta\theta = \arg\left(\frac{\mathcal{O}_2}{\mathcal{O}_1}\right) + \arg\left(\frac{\mathcal{O}_3}{\mathcal{O}_2}\right) + \arg\left(\frac{\mathcal{O}_4}{\mathcal{O}_3}\right) + \arg\left(\frac{\mathcal{O}_1}{\mathcal{O}_4}\right)\,,
\end{equation}
where each phase difference is taken in the range $(-\pi, \pi]$. The winding number $W$ is then
\begin{equation}
W = \frac{1}{2\pi} \Delta \theta\,.
\end{equation}
A non-zero integer value of $W$ may indicate the presence of a vortex (or antivortex, depending on the sign). In practice, we only perform this calculation in regions where the condensate amplitude is sufficiently large, typically where $|\mathcal{O}|/|\mathcal{O}_{\mathrm{static}}| > 0.01$, with $\mathcal{O}_{\mathrm{static}}$ being the equilibrium condensate value. In regions of very small amplitude, the phase becomes noisy and unreliable.

\begin{figure}[tbph]
    \centering
    \includegraphics[width=1\textwidth]{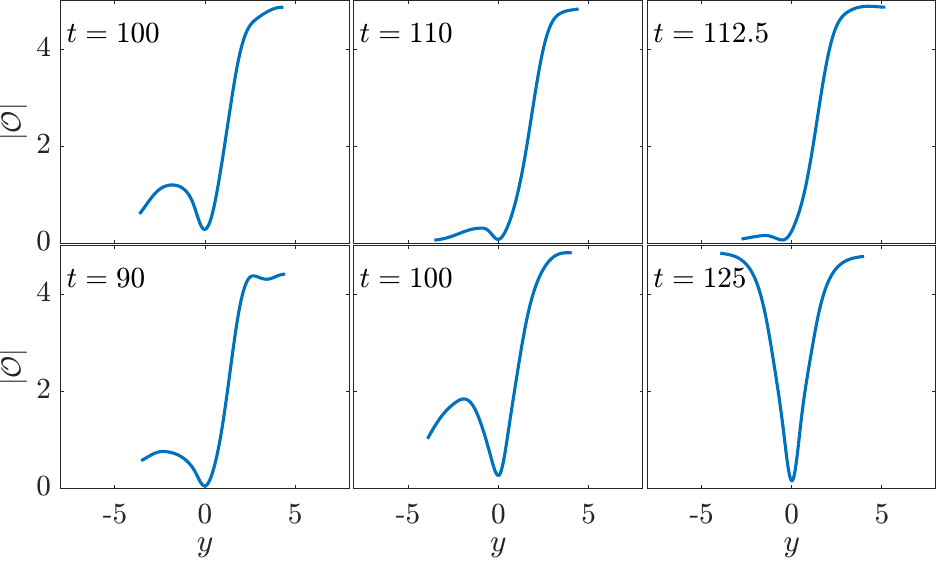}
    \caption{Time evolution of the condensate magnitude along the $y$ direction for two different collision radii at representative time slices. The origin of the horizontal axis coincides with the expected vortex formation position. \textbf{Top row}: $r_{\mathrm{co}} = 13.0078 > r_c$ (with $r_c \approx 12.9825$). A core initially forms at the origin, but the surrounding condensate subsequently decays, indicating no stable vortex. At $t = 112.5$, no winding number is found, so the profile is shown near the last detection position. \textbf{Bottom row}: $r_{\mathrm{co}} = 12.9727 < r_c$. The condensate around the core gradually grows and eventually develops into a stable vortex configuration. The data are fitted using a spline.}
\label{doublev_3conden_slice}
\end{figure}

Nevertheless, the phase winding alone is not sufficient to confirm the existence of a stable vortex, as transient phase defects can appear without a persistent amplitude dip. Therefore, we also monitor the time evolution of the condensate magnitude near the candidate vortex core. A genuine stable vortex should exhibit a sustained suppression of the condensate amplitude at the core, which gradually recovers to a finite value but remains lower than the surrounding superfluid background.

Figure~\ref{doublev_3conden_slice} illustrates this criterion for two different collision radii, $r_{\mathrm{co}} = 13.0078 > r_c$ (first row) and $r_{\mathrm{co}} = 12.9727 < r_c$ (second row), where $r_c \approx 12.9825$ is the critical radius identified in the main text. The horizontal axis origin is chosen to coincide with the position where a vortex is expected to form. In the supercritical case (first row), a core with non-zero winding initially appears, but the condensate amplitude around it decays after an initial increase, indicating that no stable vortex survives. In contrast, for the subcritical case (second row), the condensate amplitude around the core gradually grows and eventually develops into a stable vortex configuration. The data in Fig.~\ref{doublev_3conden_slice} are fitted using a spline for smoothness.

By combining the phase winding analysis with the temporal behavior of the condensate magnitude, we reliably distinguish genuine stable vortices from transient topological fluctuations. This combined method is applied throughout the multi-bubble collision simulations presented in Section~\ref{defect_generation}.

\bibliographystyle{JHEP}
\bibliography{references}

\end{document}